\documentclass[aps,prl,twocolumn]{revtex4}
\usepackage{graphicx}
\usepackage{indentfirst}
\usepackage{braket}
\usepackage{float}
\usepackage{CJK}
\usepackage{esint}
\usepackage{color}
\usepackage{subfigure}
\usepackage{amsfonts}
\makeatletter
\newcommand*{\rom}[1]{\expandafter\@slowromancap\romannumeral #1@}
\makeatother

\def\be{\begin{equation}}
\def\ee{\end{equation}}
\def\ba{\begin{eqnarray}}
\def\ea{\end{eqnarray}}

\begin{document}

\title{Cellular Automata Based Model for Pedestrian Dynamics}
\author{Quntao Zhuang}
\email{zhuangquntao@gmail.com}
\affiliation{School of Physics, Peking University, 100871, Beijing, China}
\date{April 3rd, 2011}
\begin{abstract}
We  construct a two dimensional Cellular Automata based model for the description of pedestrian dynamics.
Wide range of complicated pattern formation phenomena in pedestrian dynamics are described in the model, e.g.
lane formation, jams in a counterflow and egress behavior. Mean-field solution of the densely populated case and
numerical solution of the sparsely populated case are provided. This model has the potential to describe more flow phenomena.

\end{abstract}
\maketitle

\section{Introduction}
In pedestrian dynamics, several interesting phenomena have been intensively studied: herding effect, lane formation \cite{3} and jams in a counterflow, egress behavior\cite{10}, panic situation\cite{9} etc.. Currently two kinds of models and their extensions \cite{7,8} have been proposed to describe these phenomena:
"social force" model \cite{3} and Cellular Automata based "floor field" model \cite{4,5}. The first one follows from molecular dynamics model, it involves a group of particles interacting by long range "social-force" induced by behaviors of the individuals. However, the choice of the particular form of long-range "social force" is somehow arbitrary and generally pedestrians do not decide their behaviour in such a sophisticated way. The second one introduced certain kinds of particles that transfer inter-person forces. The particle density field is often referred to as the "floor field". This model is derived from models for insects behavior with chemotaxis. This model also achieved much success in describing the above phenomena, but is somehow a bit far from the reality.

In this paper, we present a Cellular Automata based model focusing more on how the complex patterns and interesting phenomena can emerge from the very simplicity of the person-person and person-environment interaction. (From now on, we abbreviate "Cellular Automata" to "CA".) This CA model can be regarded as an extension of the Ising Model for magnetic spins. Mean-field solution for the densely populated case of this model is discussed and numerical simulations are carried out for the less populated case. Lane formation and jams in a counterflow, as well as the egress behaviour are discussed.

The rest of the paper is organized as follows: In section \rom{2} the basic CA model setting is presented, a model Hamiltonian is given for the system; In section \rom{3}, the extended Glauber dynamics\cite{1} that governs the time evolution of the system is presented; In section \rom{4}, the dense populated case is discussed, we apply mean-field approximation to discuss the equilibrium flow. The results are in consistency with numerical results; In section \rom{5}, the sparsely populated case is discussed. We numerically simulated lane formation and jams phenomena in counterflow situation as well as egress phenomenon. Good consistency is found with others' results as well as the reality. Conclusions are given in section \rom{6}. In the appendix we present the specific derivation of the mean-field solution.

\section{Model}
Our model is based on the local decision making process of a single pedestrian, given its surrounding environment and destination direction. The surrounding environment contains the neighbouring pedestrian and their walking directions, i.e. pedestrian tend to move in the same direction with neighbours; The surrounding environment also contains the available empty spaces: pedestrian tend to avoid crowded area so they move to emptier place; they avoid walking towards a wall. The influence of the destination direction is modeled by an external field that guides them to certain directions.

We consider a group of pedestrians in certain plain space with different destination directions. We call our model "$N$ dimensional" if at most a single pedestrian can have $q=2N$ neighbouring pedestrians(see Fig \ref{config}). For simplicity we do not consider specific details of a single pedestrian, e.g. his weight, height, age, gender. etc.. Since here the space that one pedestrian occupies is the smallest space unit we are considering, the space is considered to be $n$ discrete lattices, each cell can contain at most one pedestrian. Then the total carrying capacity of the space is $n$. The number of pedestrian in this space can vary from zero up to $n$, characterizing the level of crowdedness of the space. In such a N-dimensional model, we characterize a pedestrian by q-dimensional vectors $\vec{s}^{(q)}$ and $\vec{p}^{(q)}$: $\vec{s}^{(q)}$ denotes the direction of one pedestrian and $\vec{p}^{(q)}$ denotes the condition of neighbors of one pedestrian. Since our model share similarity with Ising spin model, from now on we call a pedestrian a spin. Notice that walls and obstacles are equivalent to pedestrians that never move.

The neighbors of one lattice is coded in an order that neighbor $i$ and neighbor $i+N$ are on opposite direction to this lattice, which then forms a q-dimensional coordinates. This coordinates system distinguishes with N-dimensional conventional coordinates and is designed to describe the conditions of the $q$ neighbors of each lattice.
Considering all the possible states of a pedestrian's movement, the vectors $\vec{s}^{(q)}$ form a set denoted by $
\textbf{S}^{(q)}=\{\vec{e_0}^{(q)}\equiv(0,0,0,\dots,0),\vec{e_1}^{(q)}\equiv(1,0,0,\dots,0)
,\cdots, \vec{e}^{(i)}_{q}\equiv(0,\cdots,1,\cdots,0),\cdots,\vec{e}^{(q)}_{q}\equiv(0,0,0,\dots,1)\}
$. If $\vec{s}^{(q)}=\vec{e}^{(q)}_i(i=1,2,3\dots,q)$, then this spin's direction is towards neighbor $i$, i.e. it will move to neighbor $i$ if neighbor i is not occupied by a pedestrian. If  $\vec{s}^{(q)}=\vec{e_0}^{(q)}$ , then this lattice is empty.

Vectors $\vec{p}^{(q)}$ form a set $\textbf{P}^{(q)}=\{(p_1,p_2,\dots,p_i,\dots,p_q)|p_i\in\{1,0\},i=1,2,3,\dots,q\}$. If $p_i(i=1,2,3,\dots,q)$ equals one, then neighbor $i$ is occupied, otherwise this neighbor is empty.
Note that only by using q-dimensional coordinates can we  distinguish between these conditions:
\begin{figure}
\centering
\includegraphics[width=0.25\textwidth]{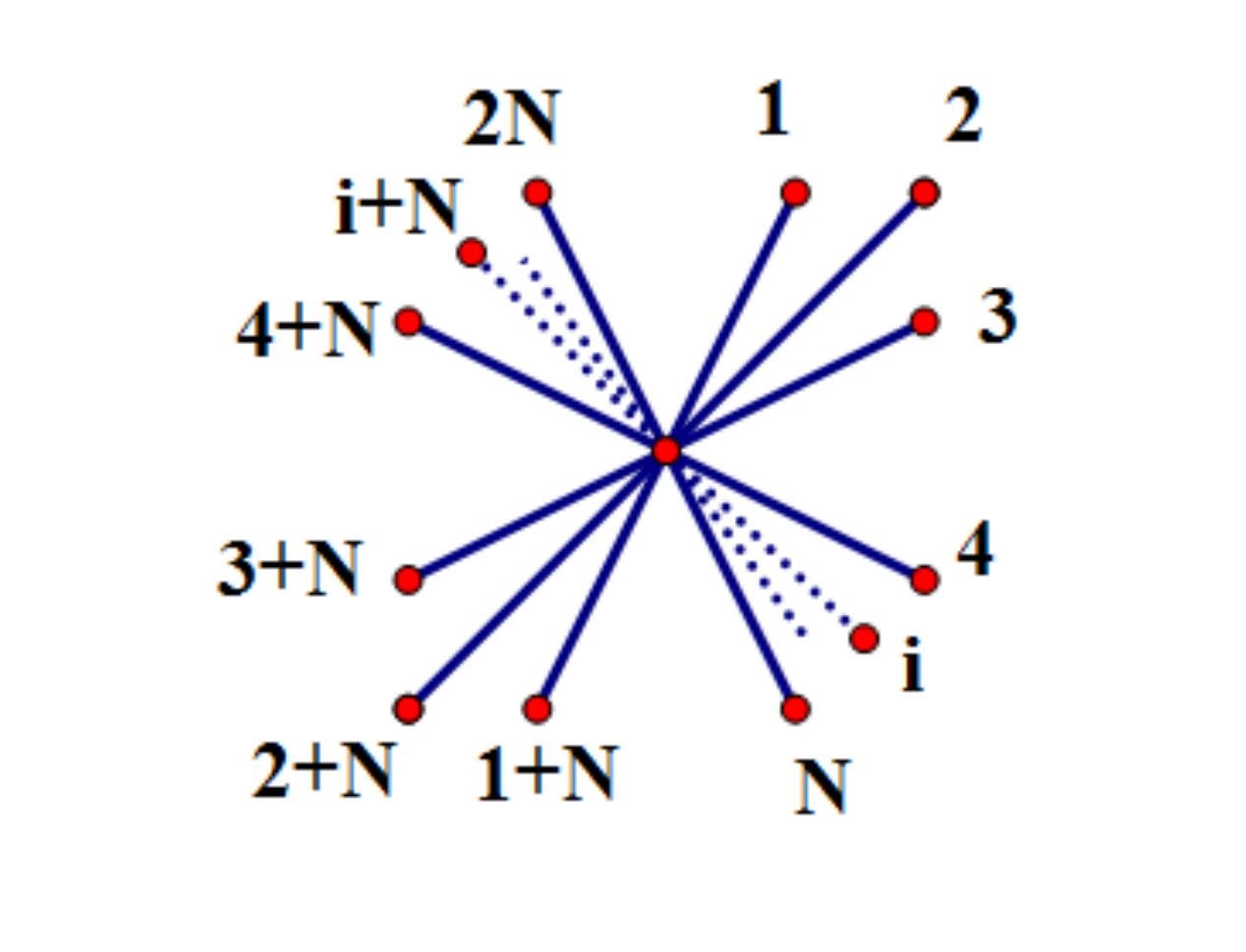}
\caption{schematic drawing of a pedestrian and its surrounding lattices of the system, the surrounding lattices are numbered so that opposite neighbours are $i$ and $i+N, i=1,\cdots,N$.
\label{config}
}
\end{figure}
\begin{figure*}
\includegraphics[width=\textwidth]{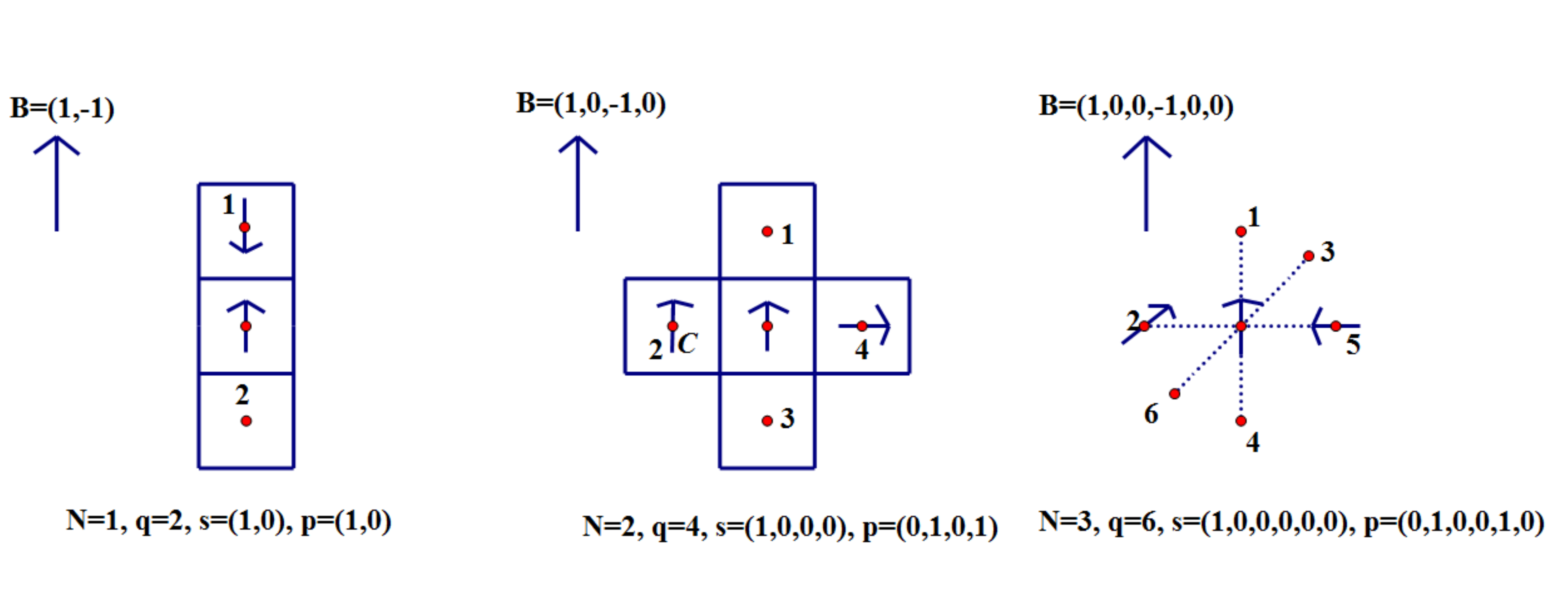}
\caption{schematic drawing of the system, $N=1,2,3$
}
\end{figure*}

\begin{itemize}
\item Neighbor $i$ is occupied, i.e. $p_i=1$, $p_{i+N}=0$.
\item Neighbor $i+N$(opposite to neighbor $i$) is occupied, i.e. $p_i=0$, $p_{i+N}=1$.
\item Both neighbor $i$ and $i+N$ are occupied, i.e. $p_i=1$, $p_{i+N}=1$.
\item Both neighbor $i$ and $i+N$ are empty, i.e. $p_i=0$, $p_{i+N}=0$.
\end{itemize}
If the total number of pedestrians in the space is $M$, since we apply periodic boundary or the space is large enough, then in all the n lattices, the number of occupied lattices is $M$ and remains a constant in all process.

In our model, time is also considered to be discrete. It is measured by steps of pedestrians. Again we do not consider the details of the difference in footstep length and pace of pedestrians. Each step is one time unit, then we can introduce the velocity direction vector of one spin $\vec{v}^{(N)}=(s_1-s_{1+N},s_2-s_{2+N},\dots,s_N-s_{2N})$ to describe the movement of one spin.

The destination direction is simulated by an external field, denoted by $\vec{B}^{(N)}=(B_1,B_2,B_3,\dots,B_{N})$ in N-dimensional coordinates. For latter use we also introduce a  q-dimensional form of the external field $\vec{B}^{(q)}=(B_1,B_2,B_3,\dots,B_{q})$ by defining $(B_{N+1},B_{N+2},B_{N+3},\dots,B_{2N})=-(B_1,B_2,B_3,\dots,B_{N})$. The interaction between neighbor spins and interaction between spins and external field is described in Hamiltonian of the system:
\be
H=-\frac{1}{2}f\sum_{<i,j>} \vec{v}^{(N)}_i\vec{v_j}^{(N)}+a\sum_{i=1}^{n} \vec{p}^{(q)}_i\vec{s}^{(q)}_i-\sum_{i=1}^{n} \vec{B}^{(q)}\vec{s}^{(q)}_i
\label{Hamiltonian_model}
\ee

The first item results from interaction between the directions of neighbor spins. Here spins have a tendency to be parallel to each other, which represents a kind of "cooperation". If neighbor spins have parallel $\vec{v}^{(N)}$, then they have smaller energy. $f>0, a>0$ are constants that determines the properties of the system. This term is responsible for herding effect.

The middle term results from the interaction  of one spin with its neighbors. Here spins have a tendency to move to neighboring lattices which is not occupied. If one spin's direction vector points to a neighboring lattice which are not occupied, then $\vec{v}^{(q)}\cdot\vec{s}^{(q)}$ is 1, otherwise it's 0. This term results in the tendency to avoid crowded situation. The last item results from the external field. The spin tends to move in the direction of the external field, which simulates its destination direction.

The Hamiltonian described in the equation (\ref{Hamiltonian_model}) can be understood directly: herding effect in micro scale is represented by the item $-\frac{1}{2}f\sum_{<i,j>}\cdot \vec{v}^{(N)}_i\vec{v_j}^{(N)}$; the tendency to avoid crowded situation is expressed by the item $a\sum_{i=1}^{n} \vec{p}^{(q)}_i\vec{s}^{(q)}_i$ and an external field plays the role of pedestrian destination direction in the item $-\sum_{i=1}^{n} \vec{B}^{(q)}\vec{s}^{(q)}_i$; The neighboring lattice occupied by wall will result in huge increase of energy, then the spin will turn away from the wall immediately.

\section{Dynamics rules}

We do not need to calculate the specific microscopic interaction force in the dynamical evolution of the system. The dynamics here is that the system is approaching a energy minimum state. An extended Glauber Dynamics \cite{1} is applied, the detailed dynamical evolution rules of the CA model is as follows:
\begin{itemize}
\item
Denote $w(\vec{s}^{(q)})$ as the probability per unit time of flipping the spin to direction $\vec{s}^{(q)}=\vec{e}^{(q)}_i(i=1,2,\dots,q)$

\be w(\vec{s}^{(q)})=\frac{e^{-\beta E(\vec{e}^{(q)}_i)}}{\sum_{j=0}^{q}e^{-\beta E(\vec{e}^{(q)}_j)}}\ee

$E(\vec{e}^{(q)}_i)$is the energy of the system when $\vec{s}^{(q)}=\vec{e}^{(q)}_i(i=1,2,\dots,q)$. $\beta=\frac{1}{k_BT}$, $T$ is the effective "temperature" of the crowd, which reflects the level of fluctuation in the system. When temperature is high, the spin has higher possibility of changing directions, corresponding to an excited crowd who are willing to change directions now and then; when temperature is low, the spin has lower possibility of changing directions, corresponding to an inert crowd who tend to keep the same direction.

\item
If one spin has the direction $\vec{s}^{(q)}=\vec{e}^{(q)}_i$ and neighbor $i$ is unoccupied, then move the spin to neighbor $i$.
\item
A periodical boundary is applied, i.e. when a spin moves to the edge it will appear ont the other side of the space.
\end{itemize}

We adopt a rule called Q2R Rule \cite{2} developed by Vichniac, basically the rule states that any Cellular Automata that changes the state of cells all at the same time can not reach equilibrium state, which results from energy conservation law. Under this dynamics, We  have the number of spins conserved. Numerical realization of the dynamical evolution share resemblance with Monte-Carlo simulation used in various areas of physics as well as math. The random number code is from ref\cite{13}.

In the following sections we will focus on two cases of the system. When the lattice is densely populated, i.e. vacancy of the lattice only appears occasionally, then the homogeneous distribution of spins allow us to apply mean-field approximation to probe its dynamical properties. When the lattice is sparsely populated, then mean-field theory no longer applies, we turn to numerical simulation to study counterflow and egress phenomena. We show that even though the dynamical rule is simple, the phenomena that emerge from the CA model are complex and interesting.

\section{Dense populated case: mean-field approximation Solution}

When the lattice is densely populated, mean-field approximation is effective to analyze the equilibrium flow of the system(for details see appendix).
We consider M of the n lattices are occupied by spins, define parameter $x_l$ to characterize the mean flow,
 \be  x_l=\left\{ \begin{array}{cc}
 <s_{N+l}-s_l>, l=1,2,...,N\\
 <s_{l}-s_{l-N}>, l=N+1,...,2N
 \end{array}\right.\ee
We introduce a mean-field $\vec{B}_{eff}^{(q)}$, which contains both the influence of the spin-spin interaction and the spin-external field interaction.
\be
B_{eff l}=fqx_l+B_l-a<p_l>, l=1,2,3,\dots, q
\ee
Then the mean-field free energy of the system can be given as:

\ba
F&\simeq& (\frac{1}{2}fqn\sum_{l=1}^{N}x_l^2+a<\vec{s}^{(q)}>\sum_{i=1}^{n}\vec{p}^{(q)}-na<\vec{p}^{(q)}><\vec{s}^{(q)}>)
\nonumber\\
&-&\frac{1}{\beta}ln(A_{n}^{M})-\frac{1}{\beta}Mln(\sum_{\vec{s}^{(q)}_i \in \textbf{S}\backslash \{\vec{e}_0\}} e^{\beta \vec{B}_{eff}^{(q)}\vec{s}^{(q)}_i})
\label{free_energy}
\ea

\subsection{Self-consistent Equations}

\begin{figure}
\subfigure[]{
\label{fig:subfig:a}
\includegraphics[width=0.5\textwidth]{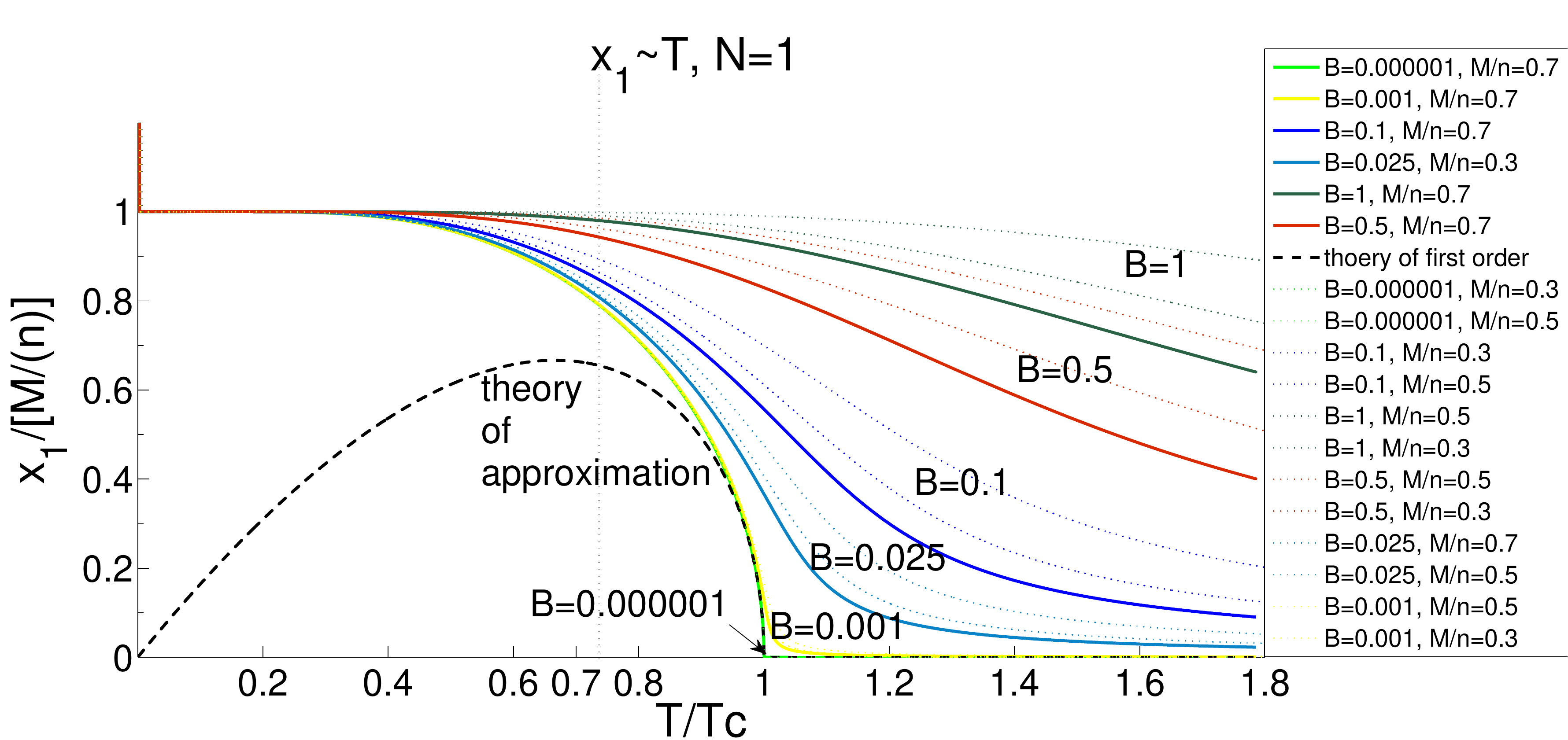}}
\subfigure[]{
\label{fig:subfig:a}
\includegraphics[width=0.5\textwidth]{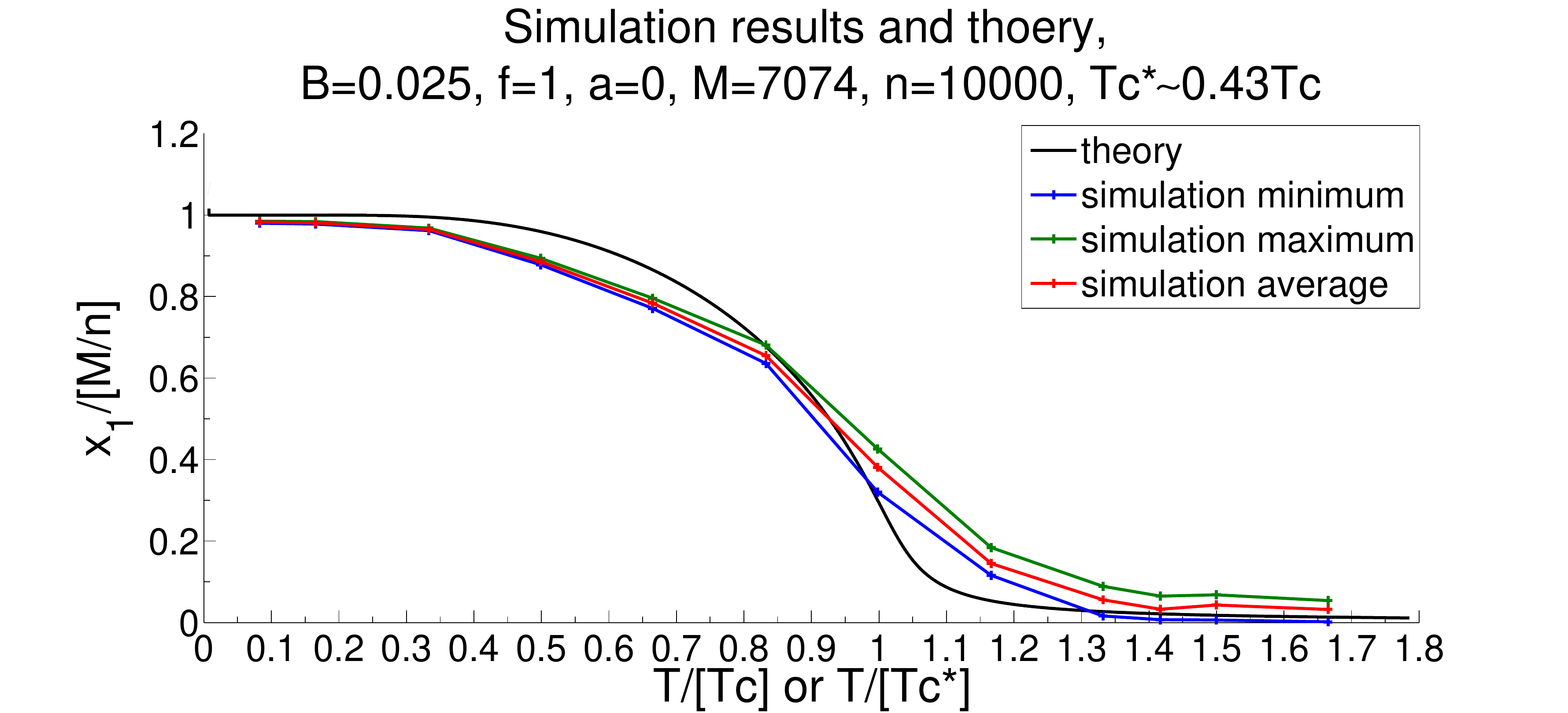}}
\subfigure[]{
\label{fig:subfig:a}
\includegraphics[width=0.5\textwidth]{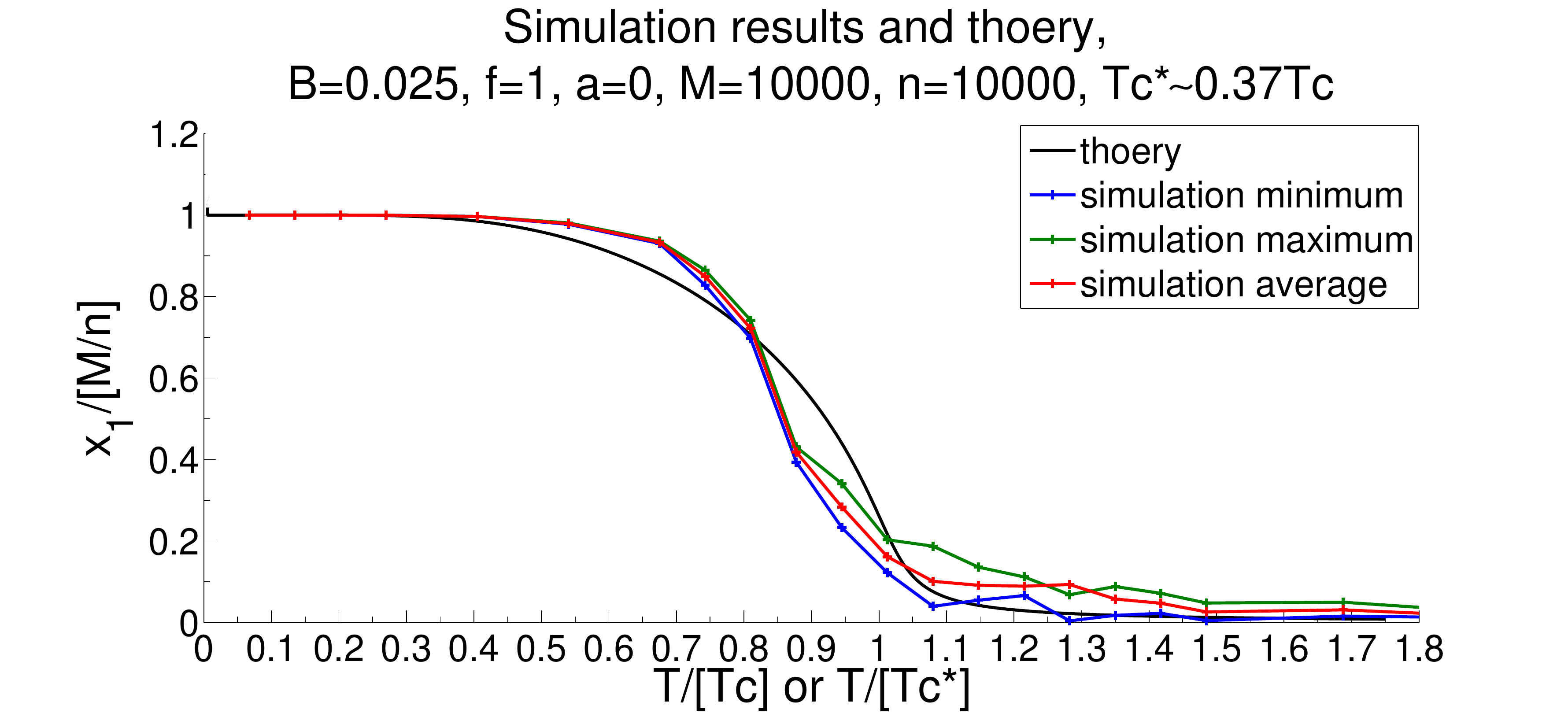}}
\caption{
(a)Numerical solution of $x_1$ in Equation (\ref{x1}), $N=1$,
$T_c\equiv\frac{2Mf}{k_Bn}$
(b)Two dimensional($N=2$) Simulation results compared with theory, critical temperature turns out to be $T_c^*\sim0.4T_c$.
(c)Two dimensional($N=2$) Simulation results compared with theory, critical temperature turns out to be $T_c^*\sim0.4T_c$. All data are averaged from over 10 separate simulations, with maximum and minimum data also plotted.
}
\label{consistent}
\end{figure}

If the system is in equilibrium, the average of $\sum_{i=1}^{n}\vec{s}^{(q)}_i$ can be calculated out:
$ \sum_{i=1}^{n}\vec{s}^{(q)}_i=-\frac{\partial F}{\partial \vec{B}_{(q)}}$
Also from the definition we have
$ \sum_{i=1}^{n}\vec{s}^{(q)}_i=n<\vec{s}^{(q)}>$. If the periodic boundary condition is adapted or the system is large enough, then We  make a further approximation $<p_l>\simeq <p_0>, l=1,2,3,\dots, q$, consequently
\be
n<s_l>=M\frac{e^{\beta (fqx_l+B_l)}}{\sum_{i=1}^{q} e^{\beta (fqx_l+B_l)}}
\ee
For simplicity and without loss of generality, we only consider the external field as follows:
$
B_1=B_0,
B_{N+1}=-B_0,
B_k=0, \mbox{    for other k}
$.
Then the solution is
\be  x_l=\left\{ \begin{array}{cc}
 0,\mbox{ }l=2,\dots,N\\
\frac{M}{n}\frac{e^{\beta(fqx_1+B_0)}-e^{-\beta(fqx_1+B_0)}}{2N-2+e^{\beta(fqx_1+B_0)}+e^{-\beta(fqx_1+B_0)}},\mbox{ }l=1
\label{x1}
 \end{array}\right.\ee
Numerical solutions of the self consistent equations are in Figure \ref{consistent}a.

When external field $B$ approaches zero there is still non-zero solution when $T$ is below certain critical point; Certain low temperature phase transition exists when $B\rightarrow 0+$. Approximation of $B \rightarrow 0+$ situation of this phase transition is done using Taylor expansion, which is showed as the black line in Figure.\ref{consistent}a.
Define the critical point as $T_c\equiv \frac{Mfq}{k_BNn}\equiv\frac{2Mf}{k_Bn}$, Taylor expansion gives:
\be
x_1= \frac{M}{n}\frac{T}{T_c}\sqrt{3(1-\frac{T}{T_c})}
\label{TC}
\ee
Since the approximation only holds when temperature is close to the critical point, We  can see that the black line fits well with the numerical solution around the critical point and deviates from the numerical solution when $T$ continues to decrease.

Two dimensional simulations are carried out to check this mean field approximation results(See Figure \ref{consistent} b.c.). For each parameters we perform 10 separate simulations and calculate the average. In order to show the variance we plot the maximum and minimum result as well. Generally the results fits well, critical temperature turns out to be $T_c^*\simeq0.4T_c$. High temperature and low temperature limits of the solution is discussed in the appendix.
\subsection{Spontaneous Self-organization}
Spontaneous self-organization exists when $B=0$, then the effective field do not contain external field.
We  require the free energy(equation (\ref{free_energy})) of the system approaches the minimum, so $\frac{dF}{d(x_l^2)}=0$.
If we  do Taylor expansion on the equation(see appendix), then the approximate solution to $\frac{dF}{d(x_l^2)}=0$ is:
\be x_l=\pm \frac{M}{Nn}\frac{T}{T_c}\sqrt{3(1-\frac{T}{T_c})},l=1,2,3,\dots, q\ee
The definition of $T_c$ is the same with equation (\ref{TC}). The uncertainty in the sign as well as in $l$ of $x_l$ is caused by fluctuations. This is essentially the symmetry breaking phenomena in our CA model. The solution is sensitive to tiny asymmetry of the initial condition. Physically it means: when temperature of the system is below $T_c$, the system will form constant flow spontaneously. Note that this spontaneous self-organization is different from the solution of the self consistent equations when $B\rightarrow 0+$: compared with the none external field condition, a tiny external field will result in great change in the system.

\section{Sparsely populated case: Counterflow and Egress phenomenon }

In a sparsely populated case, the total spins in the lattice $M$ is far smaller than the total number of lattices $n$. Then mean-field approximation does not make sense since the system is generally heterogeneous in sparsely populated case. So we turn to numerical simulation to probe the various phenomena in this case, e.g. lane formation and jams in counterflow as well as egress phenomena.

\subsection{Counterflow }
\begin{figure*}
\subfigure[]{
\label{fig:subfig:a}
\includegraphics[width=0.2\textwidth]{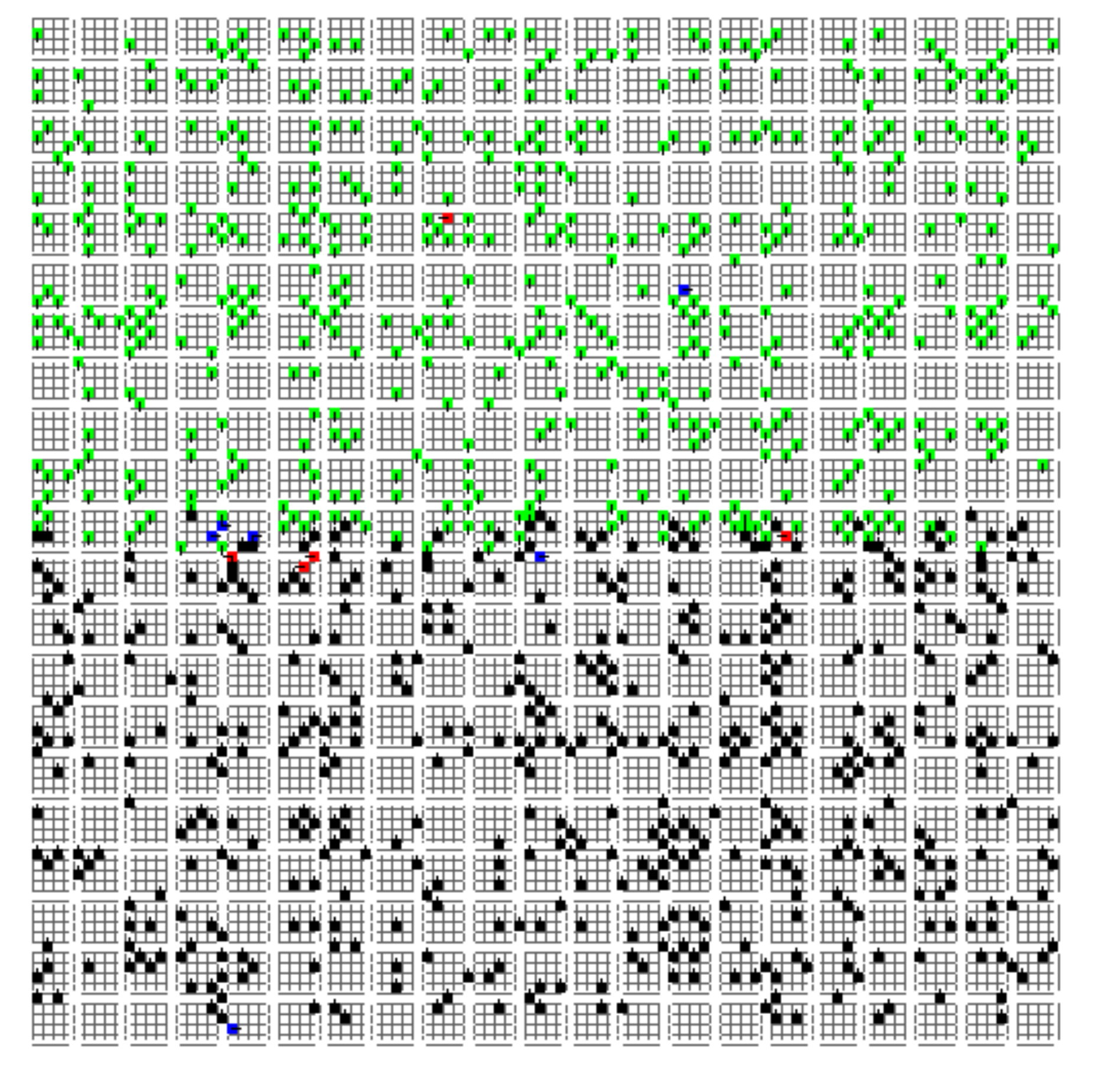}}
\subfigure[]{
\label{fig:subfig:a}
\includegraphics[width=0.2\textwidth]{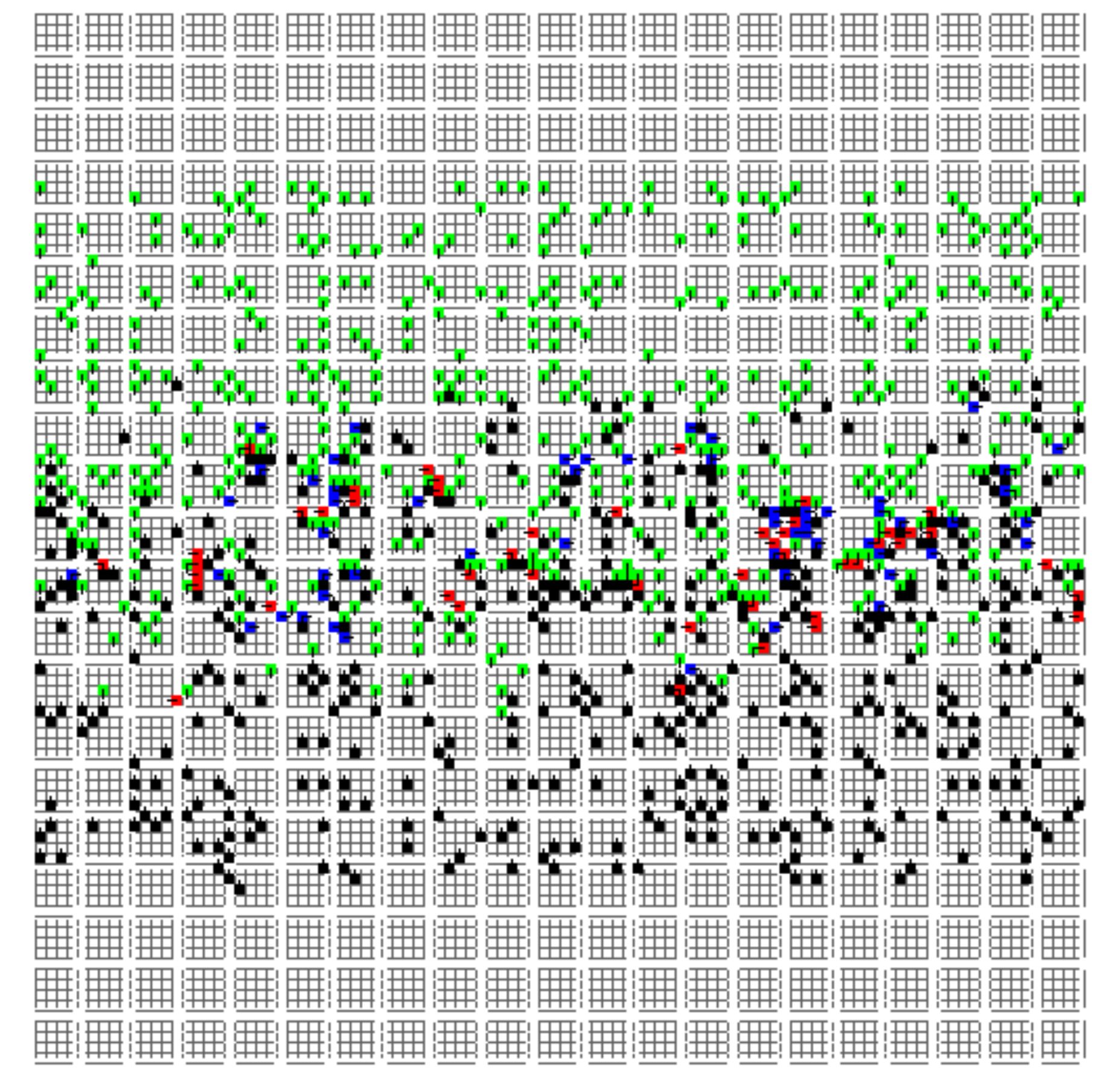}}
\subfigure[]{
\label{fig:subfig:a}
\includegraphics[width=0.2\textwidth]{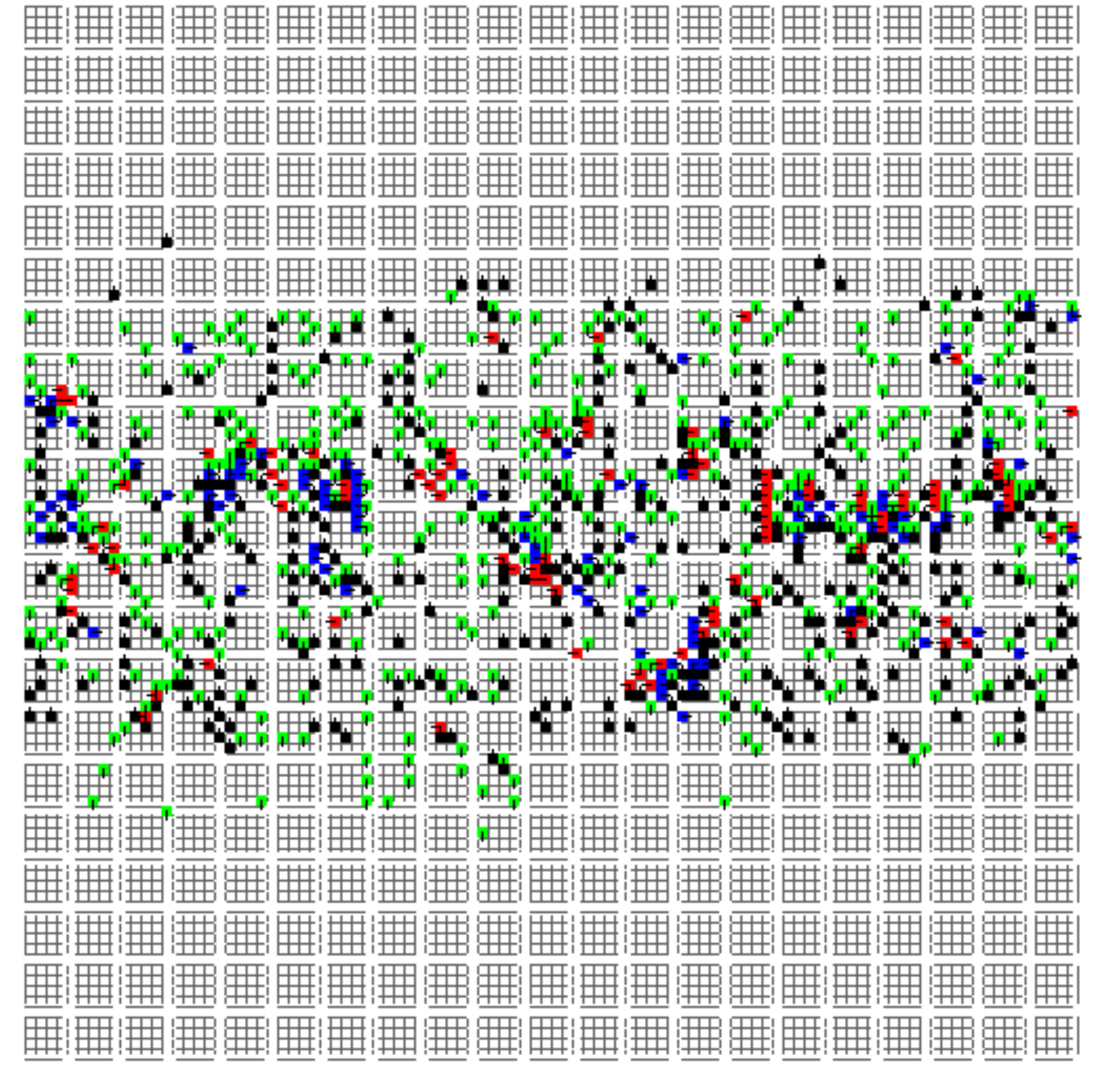}}
\subfigure[]{
\label{fig:subfig:a}
\includegraphics[width=0.2\textwidth]{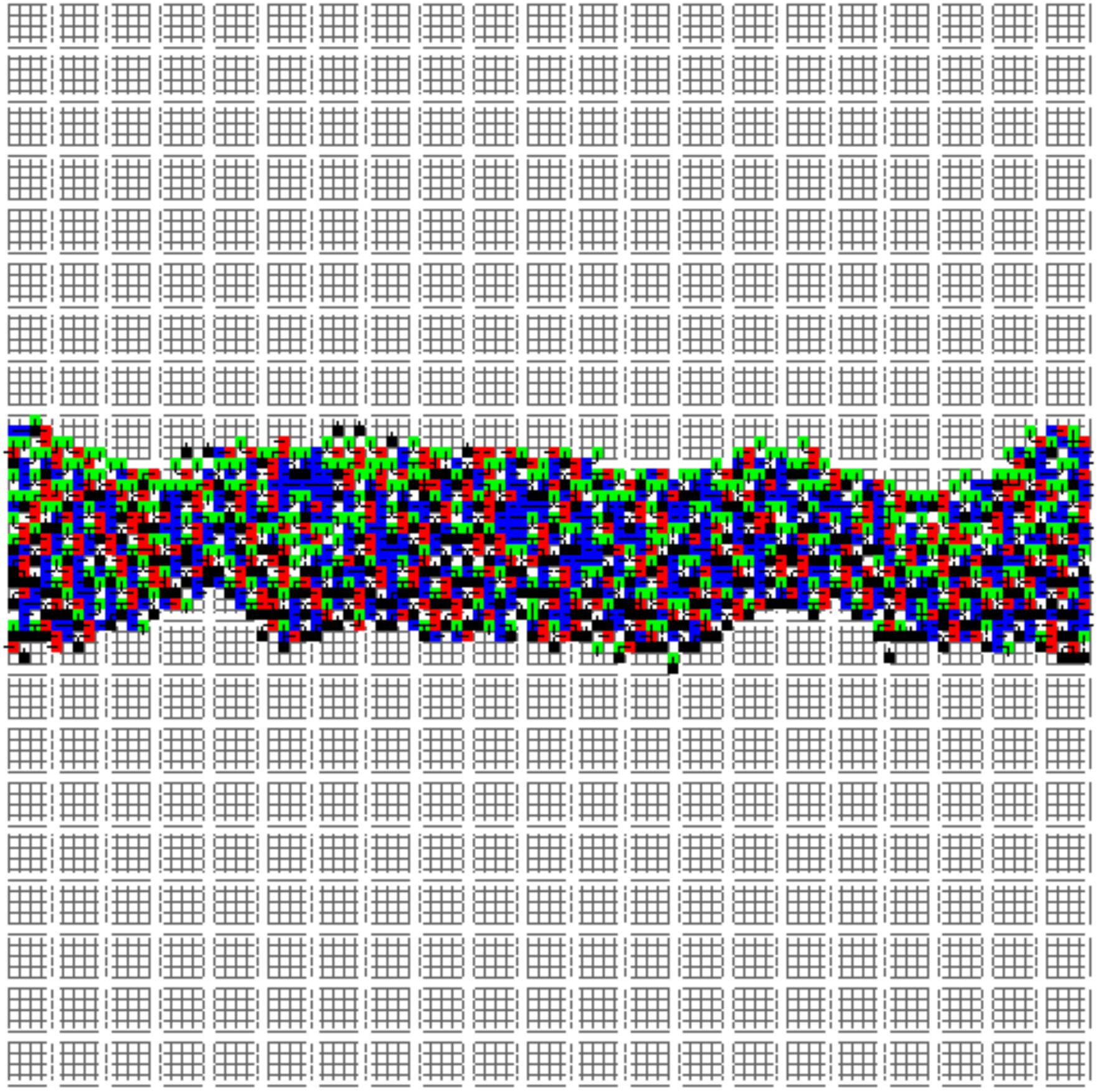}}
\caption{
color indicates the direction of pedestrian: black indicates s=(1,0,0,0)(upside), green indicates s=(0,0,1,0)(downside), blue indicates s=(0,0,0,1)(right), red indicates s=(1,0,0,0)(left).
(a)Initial state of counterflow, two crowd of pedestrian with different directions.
(b)Encounter.
(c)Merge.
(d)Jam or block phase, pedestrian tangle together and cannot form efficient flow.
}
\label{no1}
\end{figure*}
\begin{figure*}
\subfigure[]{
\label{fig:subfig:a}
\includegraphics[width=0.2\textwidth]{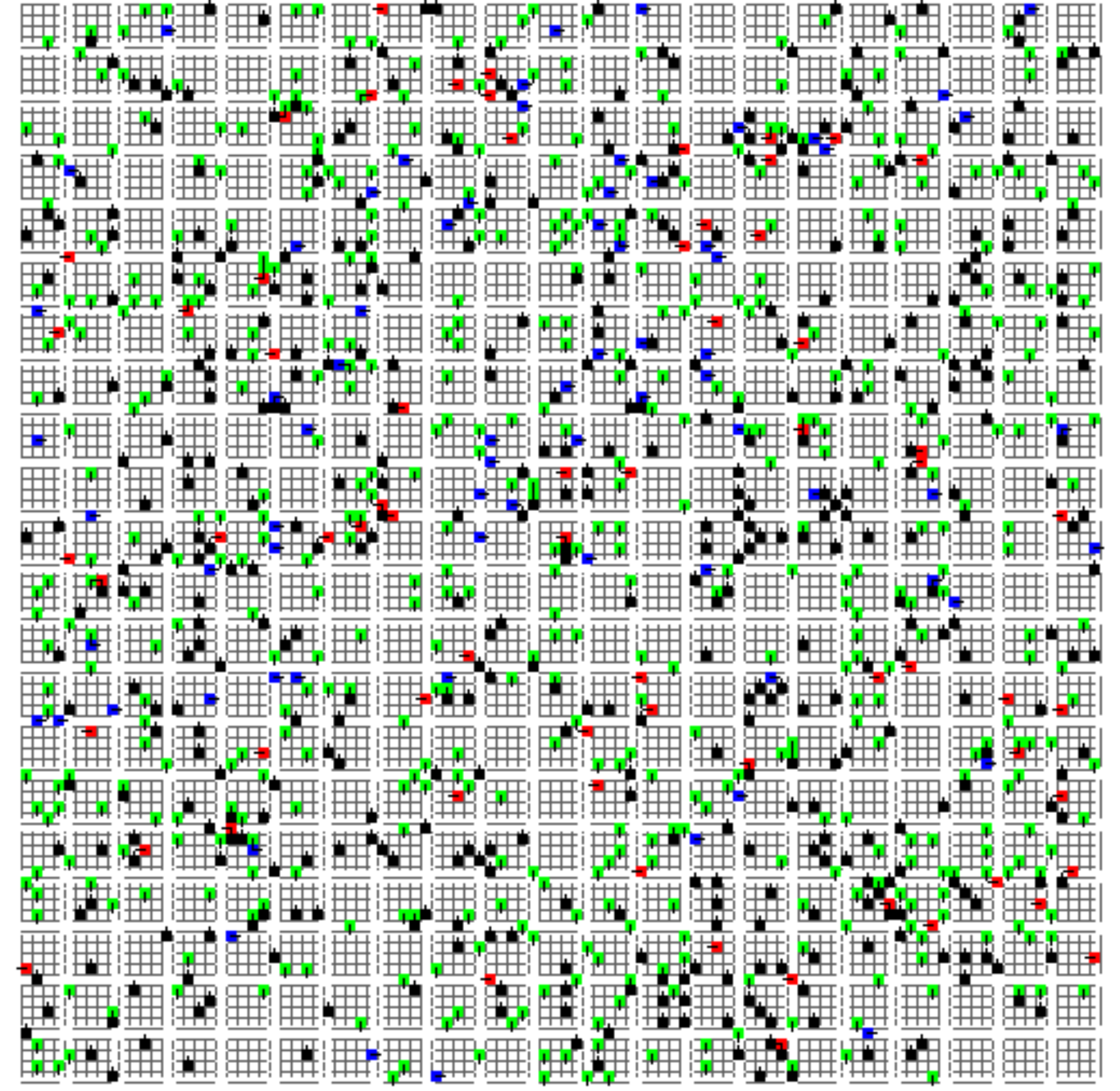}}
\subfigure[]{
\label{fig:subfig:a}
\includegraphics[width=0.2\textwidth]{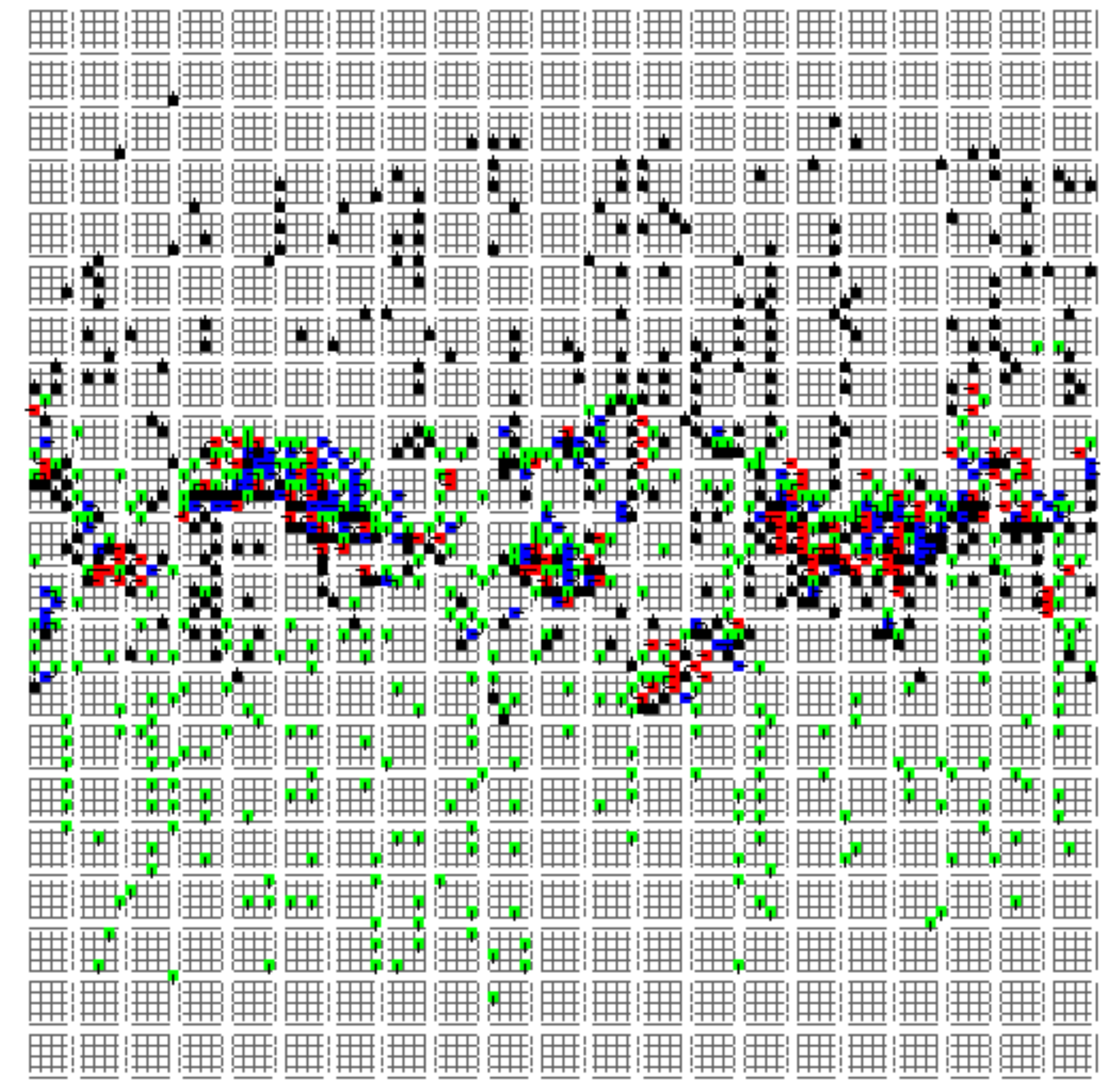}}
\subfigure[]{
\label{fig:subfig:a}
\includegraphics[width=0.2\textwidth]{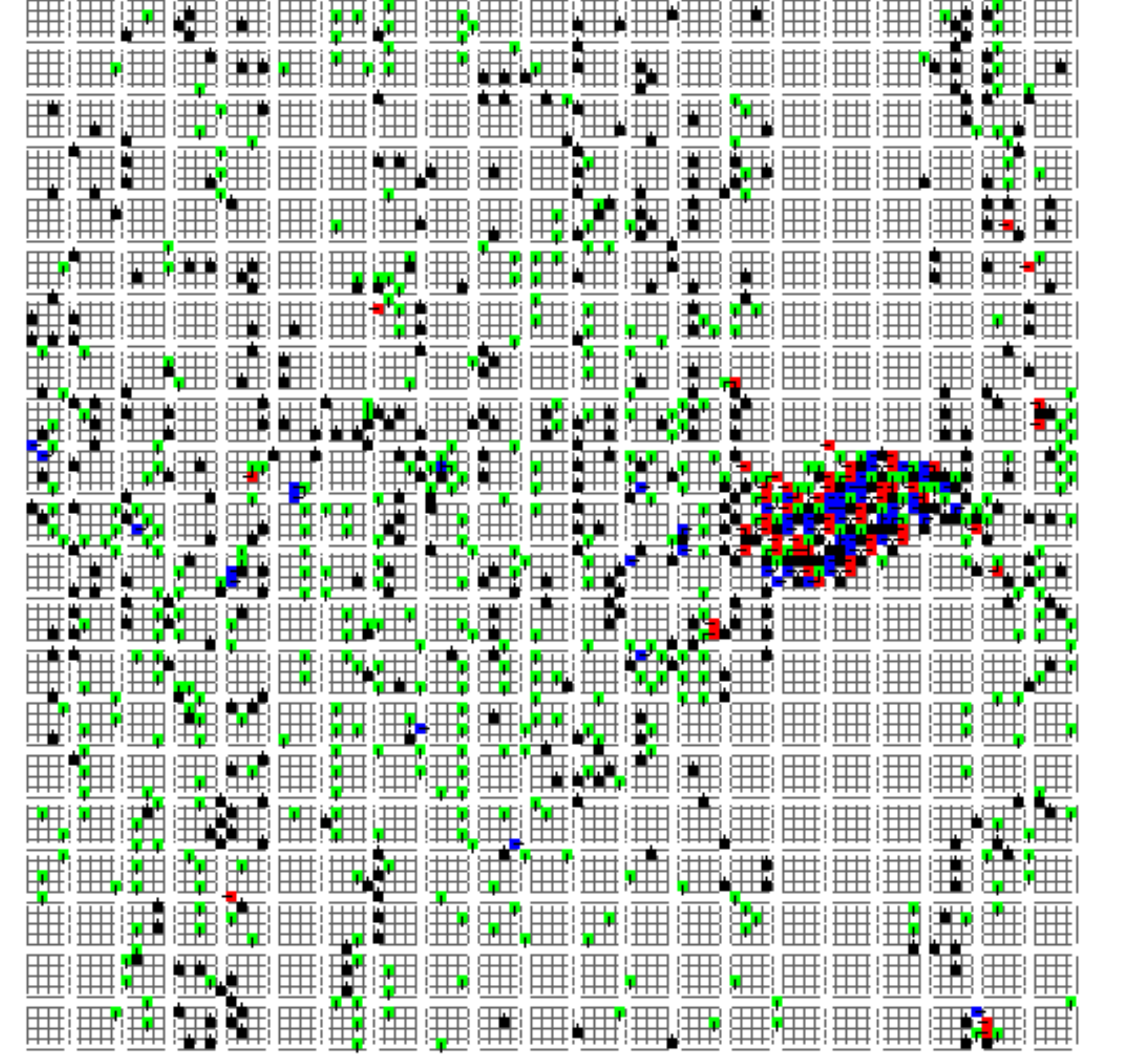}}
\subfigure[]{
\label{fig:subfig:a}
\includegraphics[width=0.2\textwidth]{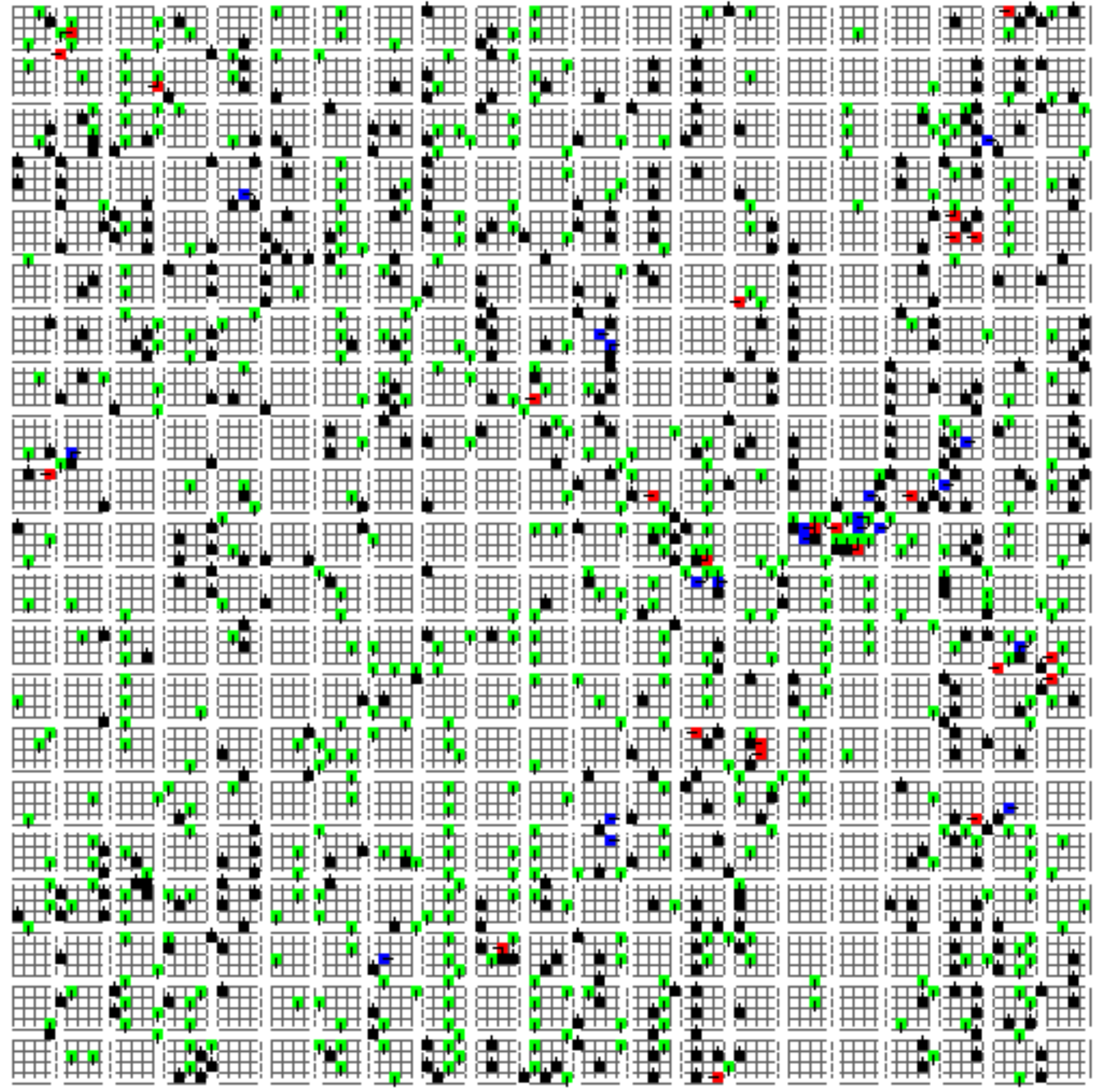}}
\subfigure[]{
\label{fig:subfig:a}
\includegraphics[width=0.2\textwidth]{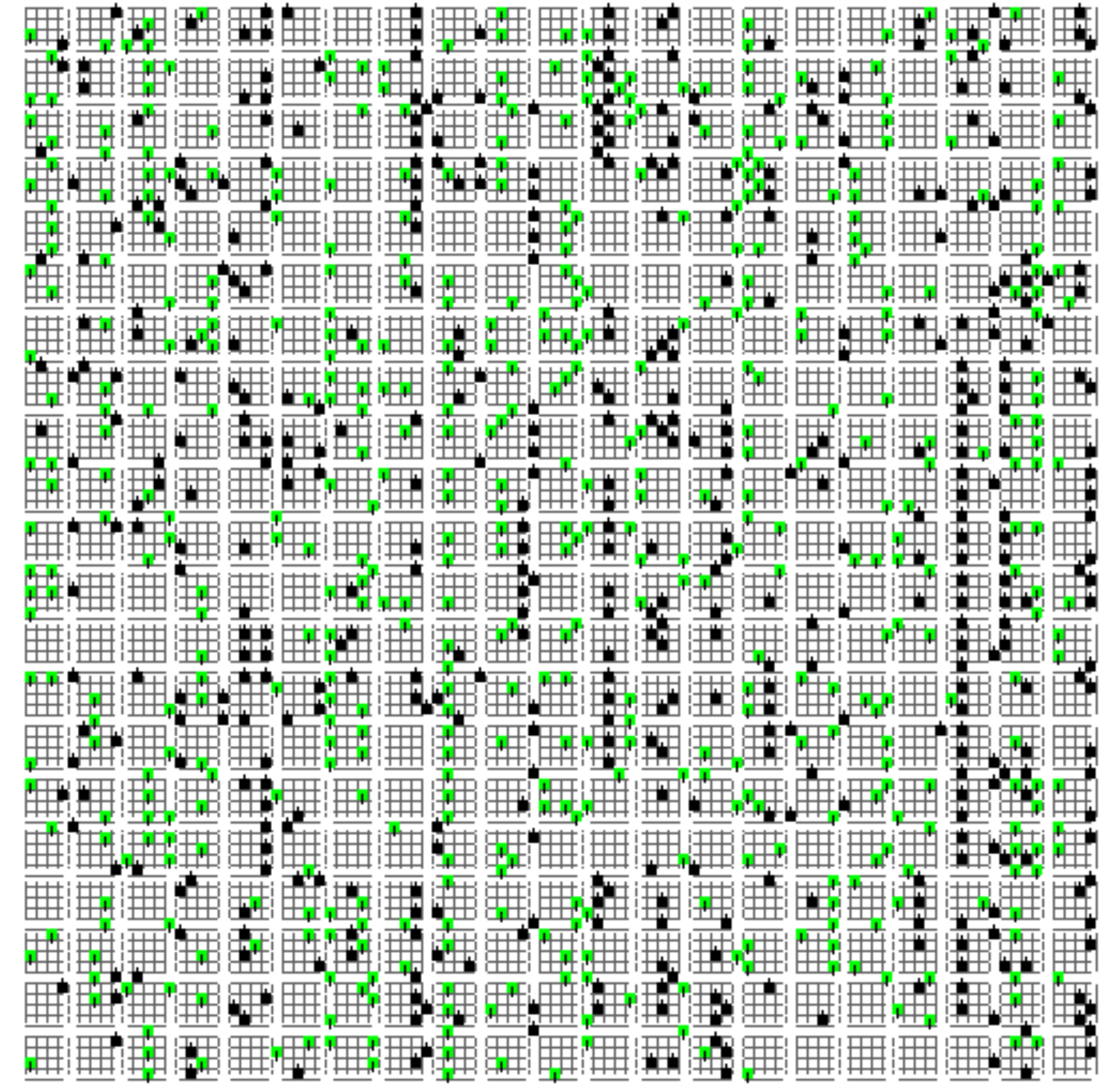}}
\subfigure[]{
\label{fig:subfig:a}
\includegraphics[width=0.2\textwidth]{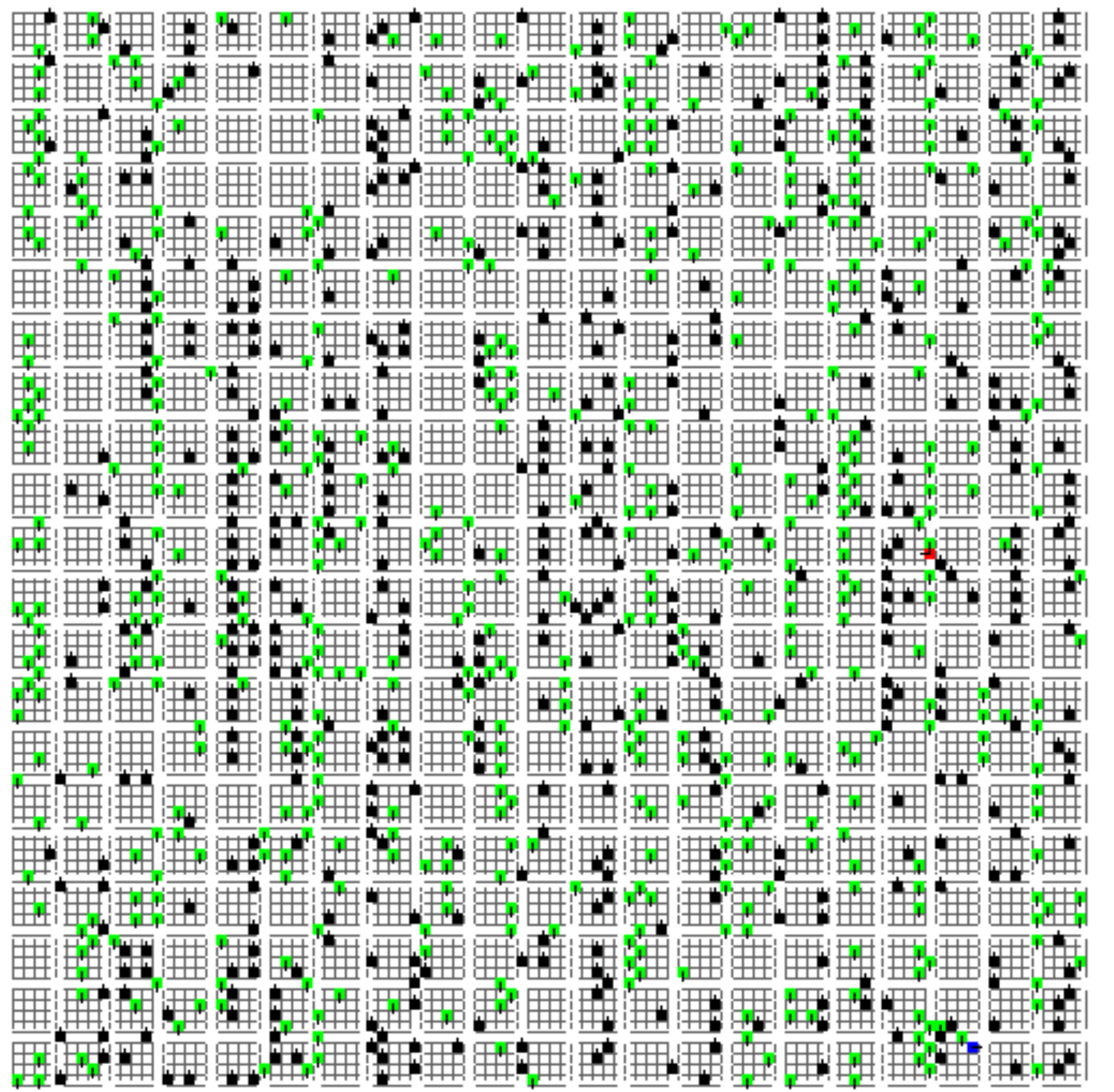}}
\subfigure[]{
\label{fig:subfig:a}
\includegraphics[width=0.4125\textwidth]{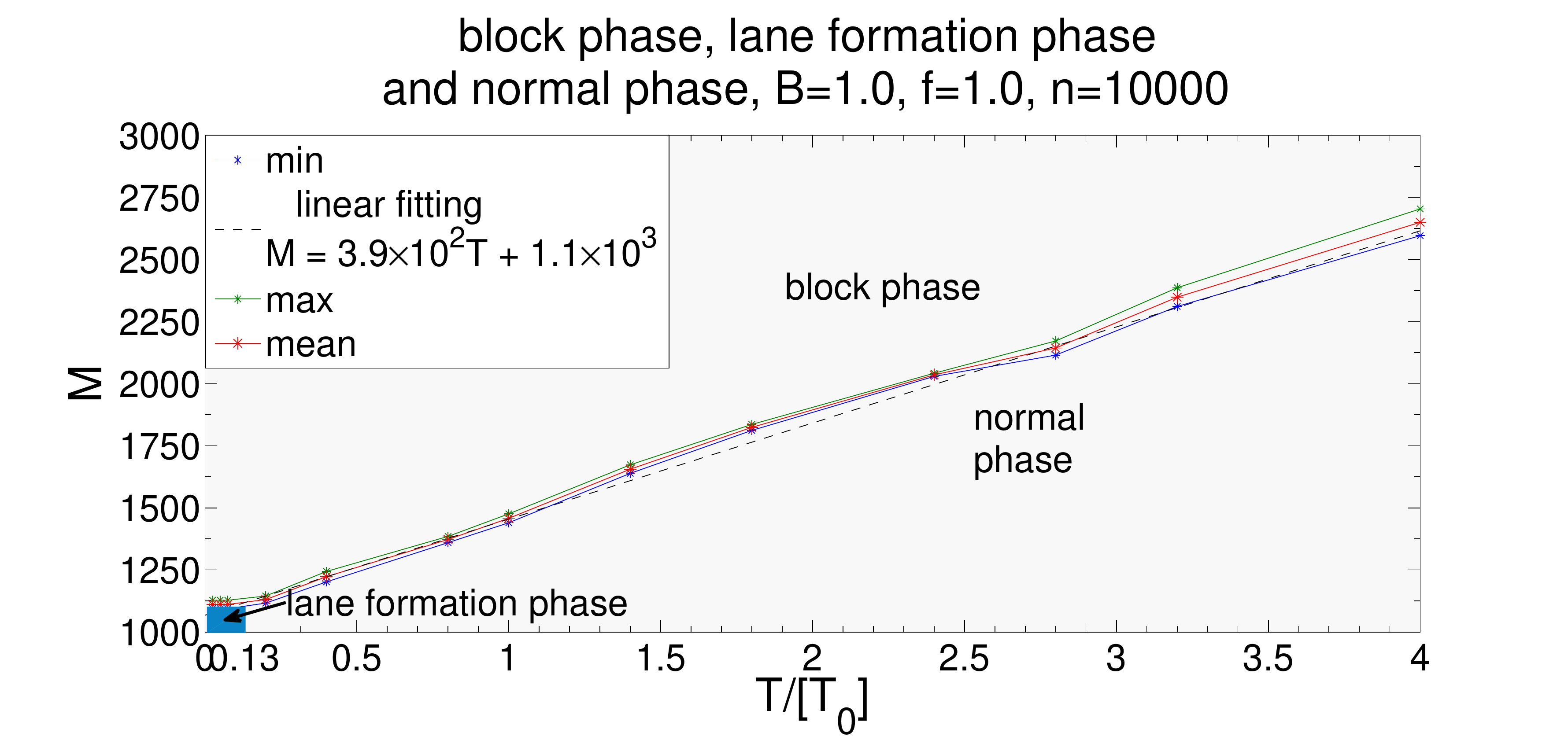}}
\caption{
color indicates the direction of pedestrian: black indicates s=(1,0,0,0)(upside), green indicates s=(0,0,1,0)(downside), blue indicates s=(0,0,0,1)(right), red indicates s=(1,0,0,0)(left). Figure (b)-(f) show the whole process of lane formation.
(a)Example of normal phase.
(b)Lane formation, transient block phenomenon before lane formation happens.
(c)Lane formation, transient block is starting to approach lane formation phase.
(d)Lane formation, transient block is disappearing.
(e)Lane formation.
(f)Lane formation.
(g)Phase diagram, $T_0$ is the numerical unit of temperature
}
\label{phase}
\end{figure*}

Study on counterflow can be found in Refs.\cite{3,4,5,7}. Basically, the counterflow situation refers to the encounter of two crowds of pedestrian coming from two opposite directions. The phenomena can take place in system with a periodical boundary or an open boundary and also in a narrow corridor. Several distinctive phenomena will happen depending on the density of the crowd, the average speed of the pedestrian and the crowd's level of anxiety. We conducted several simulations of our model and showed the different phases of the pedestrian under different parameters. Since it's counterflow, in our simulation pedestrian are labeled into two kinds, either tends to be parallel or anti-parallel with the external field. The initial positions of pedestrian are randomly chosen.

Choose the parameters as follows: $a=100, f=1,B=1, n=10000$. This corresponds to a very active way of walking, pedestrians cooperate and also they avoid crowded areas.
Figure\ref{no1}a is a snapshot of the initial condition, color indicates the direction of pedestrian: black indicates s=(1,0,0,0)(upside), green indicates s=(0,0,1,0)(downside), blue indicates s=(0,0,0,1)(right), red indicates s=(1,0,0,0)(left).

After encounter process(See Figure\ref{no1}b) and merge process(See Figure \ref{no1}c), three kinds of metastable states can be observed. We call the three states different phases of the system: "block phase"(Figure \ref{no1}d), i.e. pedestrian of different directions block each other's road and cannot move to the other side; "normal phase"(Figure \ref{phase}a), i.e. the system is not in order and fluctuates greatly; "lane formation" phase(Figure \ref{phase}e,f), i.e. the pedestrians form lanes and move efficiently. Metastable means the states keep changing in specific shapes, but remain the same kind of pattern. The final state is dominated by the total number of spins $M$ and temperature $T$, the phase diagram is shown in Figure\ref{phase}g. Lane formation phase and normal phase do not have a clear boundary, lane formation phase has a rectangle shape in the diagram approximately. A linear boundary between the block phase and normal phase is discovered.

Block phase(Figure \ref{no1}d) can also be called jam phase in consistence with the terms used by others.
In normal phase(Figure \ref{phase}a) the system is not in order and fluctuates greatly, though lane formation phase can be seen, but the lanes stay just for several time steps and are not stable(In each step of time a pedestrian can take no more than one step).
Transient block phenomenon(See Figure \ref{phase}b,c,d) is observed before lane formation. In lane formation phase(Figure \ref{phase}e,f), long lanes are formed and the system remain relatively stable. Pedestrian can move fluently and the transport efficiency is maximum.

\subsection{Egress phenomenon}

\begin{figure*}
\subfigure[]{
\label{fig:subfig:a}
\includegraphics[width=0.2\textwidth]{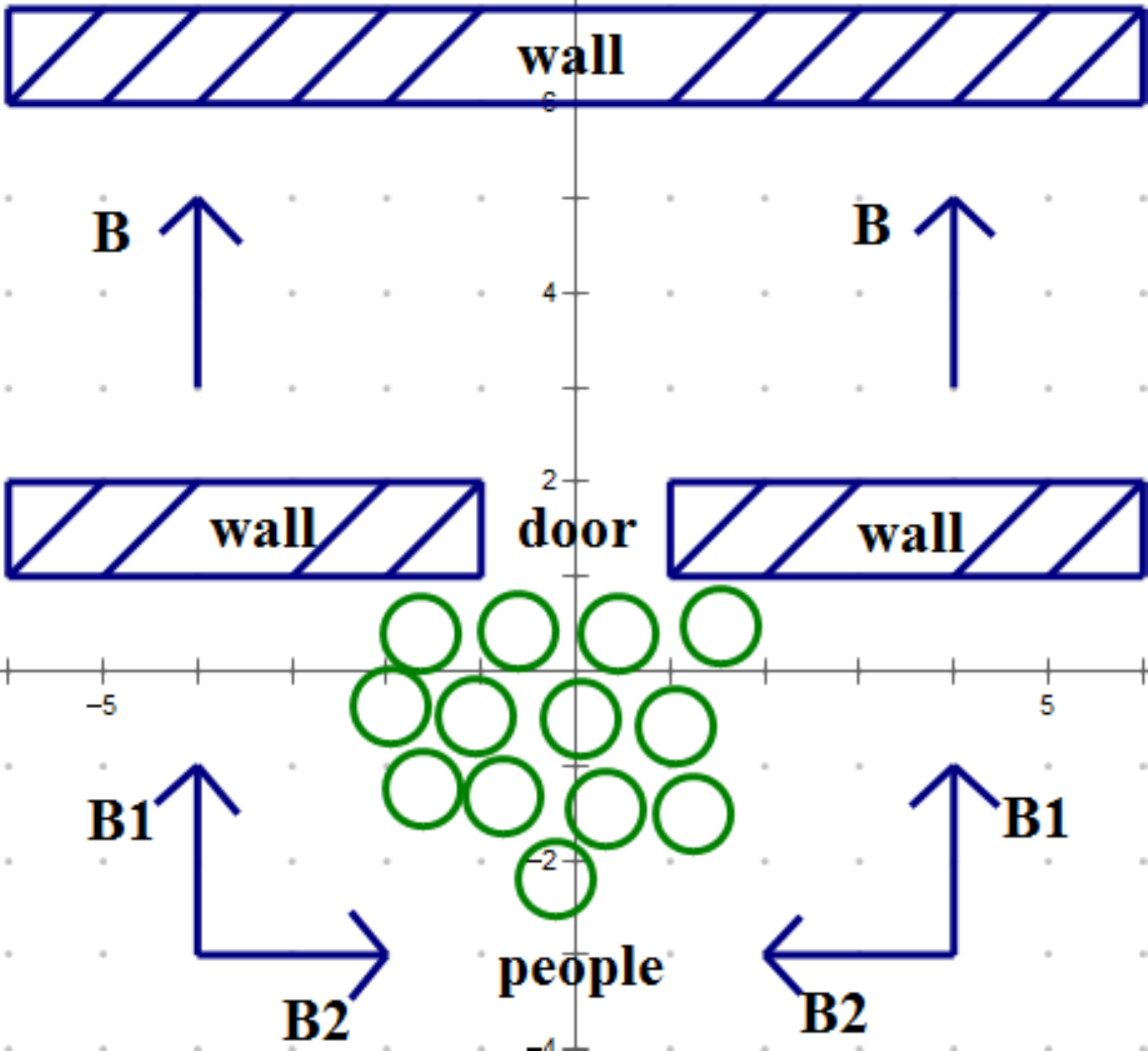}}
\subfigure[]{
\label{fig:subfig:a}
\includegraphics[width=0.2\textwidth]{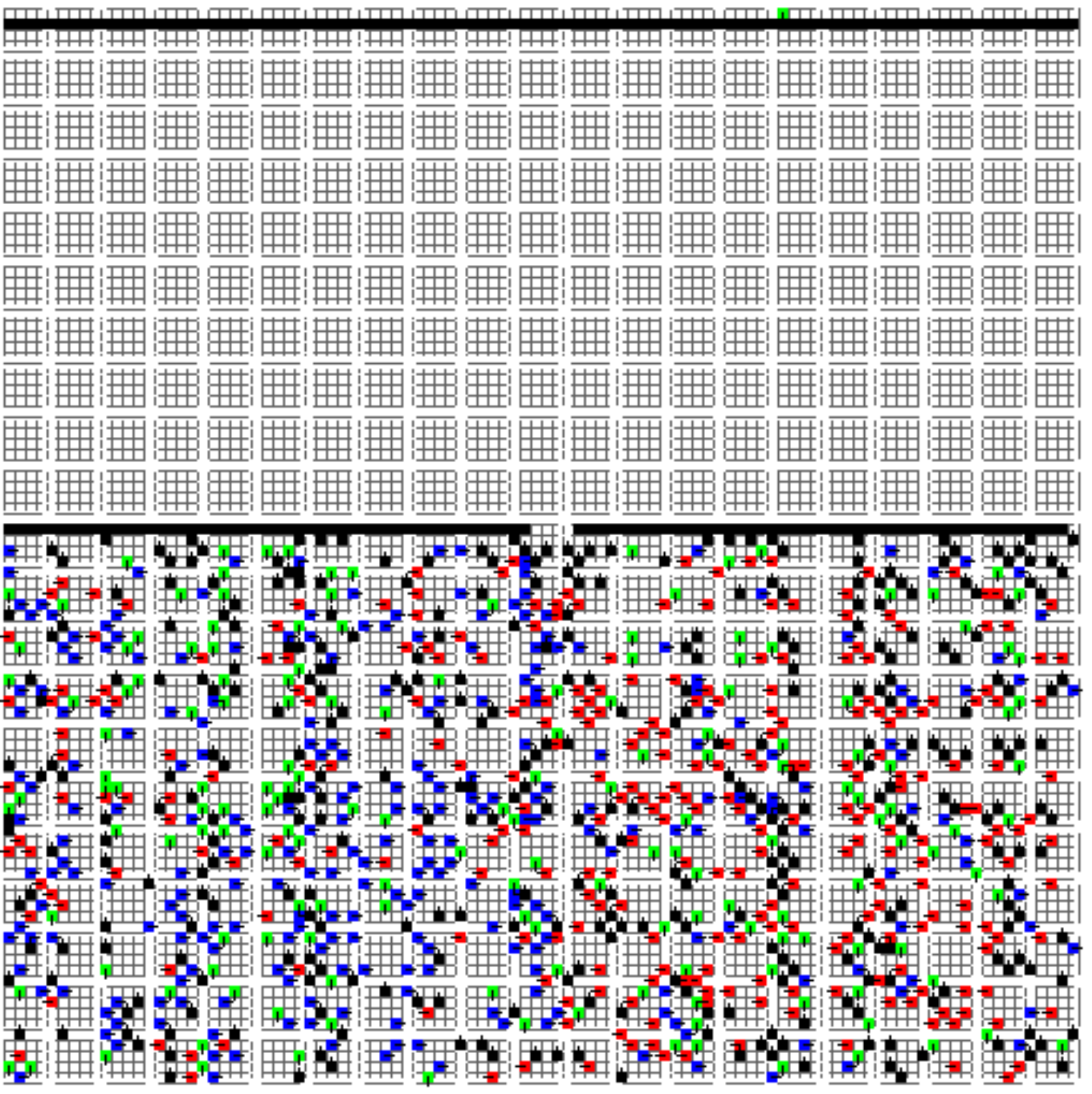}}
\subfigure[]{
\label{fig:subfig:a}
\includegraphics[width=0.2\textwidth]{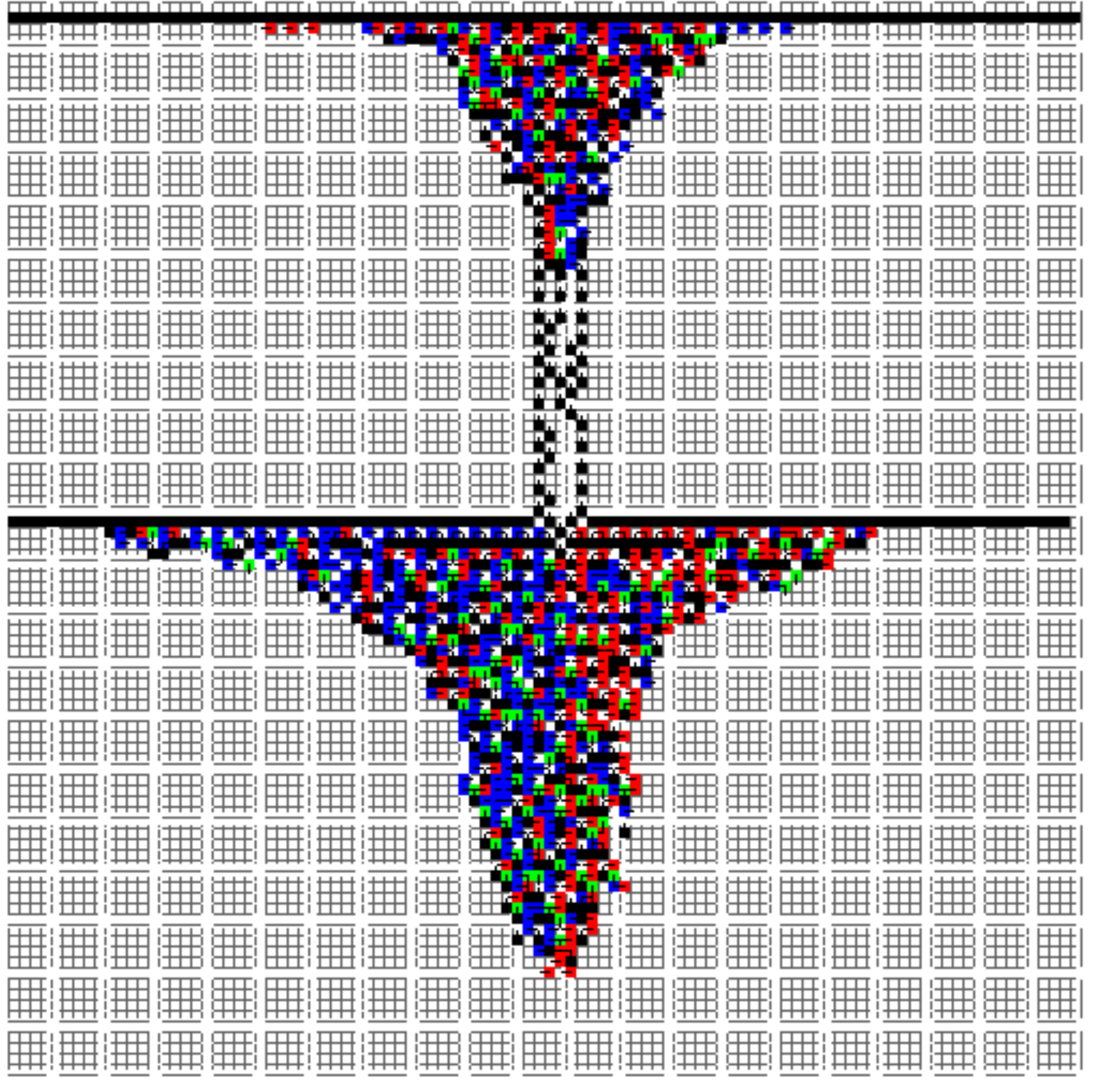}}
\subfigure[]{
\label{fig:subfig:a}
\includegraphics[width=0.2\textwidth]{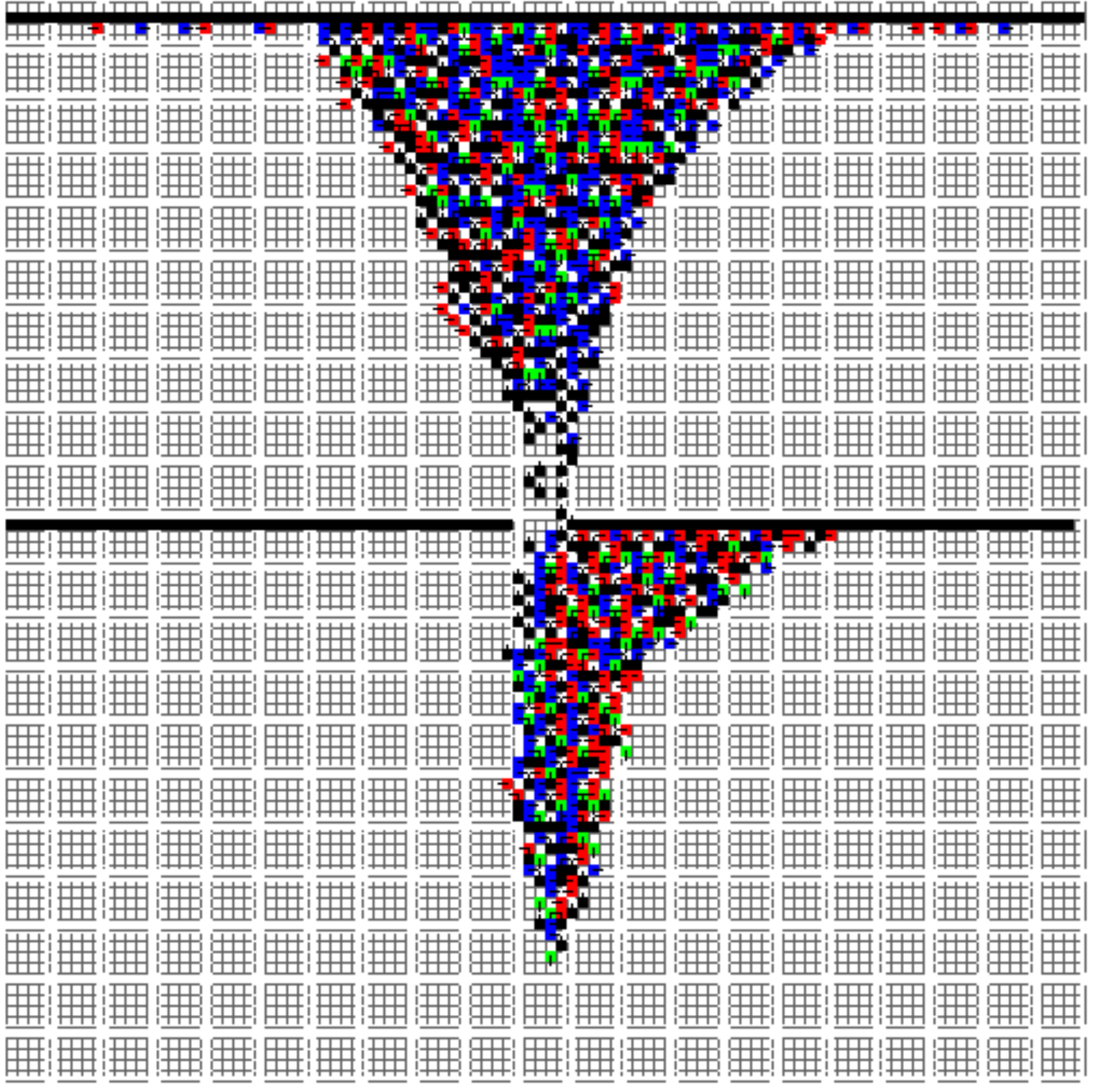}}
\caption{
(a)schematic drawing of the external field B and the situation of egress.
(b)Initial condition.
(c)Egress behavior.
(d) asymmetric egress, a lane is formed on the left side of the crowd around the door so that the left side go through the door first while the right side pedestrian remain blocked.
\label{egress}
}
\end{figure*}

\begin{figure}
\includegraphics[width=0.5\textwidth]{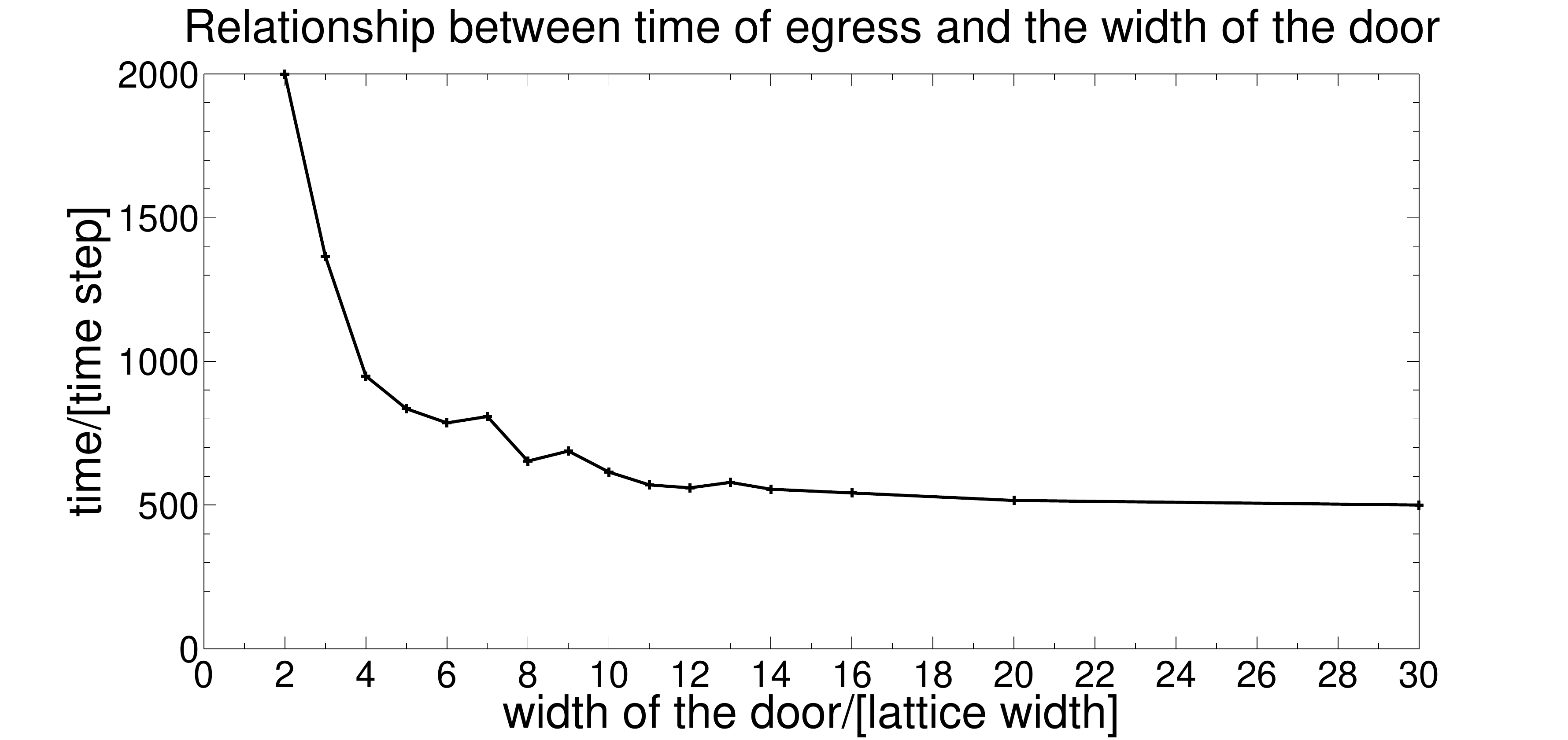}
\caption{egress time and the width of the door, critical width is between $3-4$ lattice length
}
\label{door}
\end{figure}
Egress phenomenon\cite{10} refers to the process of a group of pedestrian leaving a room in a certain period of time.
An external field acts as the aim of the crowd that guides the crowd to the door. The external field is set so that pedestrian are moving towards the up side of the space from the exit in the middle(See Figure \ref{egress}a). Then the crowd clogs around the door until the whole egress is over(See Figure \ref{egress}b,c,d). During the egress, asymmetric behavior of the clogging crowd around the exit is observed(see Figure \ref{egress}d). This phenomenon formed in the following process: a pedestrian either from the left side or the right side of the clogging crowd enter the exit, then a lane is formed since the pedestrian of his side follows his movement while the pedestrian from the other side are expelled by the lane and await until all the pedestrian from the left side enter the exit.

Figure \ref{door} shows the relationship between egress time and door width. Certain critique width is discovered, the result is consistent with Ref.\cite{10}, the critical width of a door is $3 \sim 4$ lattice width. According to Ref.\cite{5}, each lattice is approximately $40\mbox{cm}\times 40\mbox{cm}$, which should be narrowly able to contain one standing pedestrian, thus the critical width of a door is about $1.2\mbox{m}\sim 1.6\mbox{m}$, which is approximately consistent with our daily life, e.g. typical door in Peking University dorm building.

\section{Conclusion}

We construct a Cellular Automata based model to describe pedestrian dynamics. We focus on the emergence of complex phenomena resulting from simple local interaction between pedestrian.
Analytical mean-field solution of the model is given to describe dense populated case. Simulations are carried out to verify the results.
Numerical simulations are carried out to probe the sparsely populated case, which is beyond the mean-field approximation. Typical phenomena like lane formation \cite{3} and jams in a counterflow and egress behavior\cite{10} are found. Block phase, normal phase and lane formation phase are distinguished and the phase diagram shows that only in very limited situation lane formation can be observed. A linear boundary between block phase and normal phase is discovered. Asymmetric egress behavior is discovered. The critical width for the door is qualitatively measured, results are consistent with Ref.\cite{10} and real life.

Further application of this Cellular Automata based model can be expected in qualitative descriptions of other flow phenomena. It shows that Cellular Automata is a powerful tool to describe non-linear and complex phenomena.

\section{Acknowledgements}
This research was conducted by the author during his sophomore year in School of Physics, Peking University for participating in
the "Challenge Cup-National Contest of College Students' Scientific and Technological".
The author thanks Prof. Hongli Wang for helpful discussions, as well as Xun Gao for
introducing pedestrian dynamics and the notion of Cellular Automata to the author.

\begin{widetext}

\section{Appendix Derivation}
\subsection{Details of Mean-field approximation Solution}
Now we study the hamiltonian of the system

\[
H=-\frac{1}{2}\sum_{<i,j>}f\cdot \vec{v}^{(N)}_i\vec{v_j}^{(N)}+\sum_{i=1}^{n}a\cdot \vec{p}^{(q)}_i\vec{s}^{(q)}_i-\sum_{i=1}^{n} \vec{B}^{(q)}\vec{s}^{(q)}_i
\]

we have:

\[\vec{v}^{(N)}_i\vec{v_j}^{(N)}=(s_{i1}-s_{iN+1},s_{i2}-s_{iN+2},\dots,s_{iN}-s_{i2N})\cdot(s_{j1}-s_{jN+1},s_{j2}-s_{jN+2},\dots,s_{jN}-s_{j2N})
\]
\[=\sum_{k=1}^{N}s_{ik}\cdot s_{jk}+\sum_{k=1}^{N}s_{ik+N}\cdot s_{jk+N}-(\sum_{k=1}^{N}s_{ik}\cdot s_{jk+N}+\sum_{k=1}^{N}s_{ik+N}\cdot s_{jk})\]

Define \[\vec{s}^{(q)}_i\bigodot \vec{s}^{(q)}_j=\colon \sum_{k=1}^{N}s_{ik}\cdot s_{jk+N}+\sum_{k=1}^{N}s_{ik+N}\cdot s_{jk}\]

It is easy to see that:

\[\vec{s}^{(q)}_i\bigodot \vec{s}^{(q)}_j=\vec{s}^{(q)}_j\bigodot \vec{s}^{(q)}_i\] and

 \[(\vec{A}^{(q)}+\vec{B}^{(q)})\bigodot \vec{C}^{(q)}=\vec{A}^{(q)}\bigodot
\vec{C}^{(q)}+\vec{B}^{(q)}\bigodot \vec{C}^{(q)}\]

then

\[\vec{v}^{(N)}_i\vec{v_j}^{(N)}=\vec{s}^{(q)}_i\cdot \vec{s}^{(q)}_j-\vec{s}^{(q)}_i\bigodot \vec{s}^{(q)}_j\]

Apply mean-field approximation to it, we have

\[
\vec{s}^{(q)}_i\cdot \vec{s}^{(q)}_j=(\vec{s}^{(q)}_i-<\vec{s}^{(q)}_i>+\vec{s}^{(q)}_i)\cdot(\vec{s}^{(q)}_j-<\vec{s}^{(q)}_j>+<\vec{s}^{(q)}_j>)
\]
\[
\simeq (\vec{s}^{(q)}_i+\vec{s}^{(q)}_j)\cdot<\vec{s}^{(q)}_j>-<\vec{s}^{(q)}_i>\cdot<\vec{s}^{(q)}_j>
\]

since
$<\vec{s}^{(q)}_i>=<\vec{s}^{(q)}_j>$, denote them as $<\vec{s}^{(q)}>$, then

\[
\vec{s}^{(q)}_i\cdot \vec{s}^{(q)}_j
\simeq (\vec{s}^{(q)}_i+\vec{s}^{(q)}_j)\cdot<\vec{s}^{(q)}>-<\vec{s}^{(q)}>\cdot<\vec{s}^{(q)}>
\]

Similarly,
\[
\vec{s}^{(q)}_i\bigodot \vec{s}^{(q)}_j
\simeq (\vec{s}^{(q)}_i+\vec{s}^{(q)}_j)\bigodot<\vec{s}^{(q)}>-<\vec{s}^{(q)}>\bigodot<\vec{s}^{(q)}>
\]

\[
\vec{p}^{(q)}_i\vec{s}^{(q)}_i
\simeq \vec{p}^{(q)}_i\cdot<\vec{s}^{(q)}>+<\vec{p}^{(q)}>\cdot\vec{s}^{(q)}_i-<\vec{p}^{(q)}>\bigodot<\vec{s}^{(q)}>
\]
Then we have the summation as follows:
\[
\sum_{<i,j>} \vec{v}^{(N)}_i\vec{v_j}^{(N)}\simeq 2q<\vec{s}^{(q)}>\sum_{i=1}^{n}\vec{s}^{(q)}_i-nq<\vec{s}^{(q)}><\vec{s}^{(q)}>-\]
\[
(2q<\vec{s}^{(q)}>\bigodot\sum_{i=1}^{n}\vec{s}^{(q)}_i-nq<\vec{s}^{(q)}>\bigodot<\vec{s}^{(q)}>)
\]

\[
\sum_{i=1}^{n} \vec{p}^{(q)}_i\vec{s}^{(q)}_i\simeq <\vec{s}^{(q)}>\sum_{i=1}^{n}\vec{p}^{(q)}_i+<\vec{p}^{(q)}>\sum_{i=1}^{n}\vec{s}^{(q)}_i-n<\vec{p}^{(q)}><\vec{s}^{(q)}>
\]

and it is easy to see
\[<\vec{s}^{(q)}>\bigodot<\vec{s}^{(q)}>-
<\vec{s}^{(q)}>\cdot<\vec{s}^{(q)}>=-\sum_{l=1}^{N}x_l^2 \mbox{, letting $x_l=(<s_l>-<s_{l+N}>)$,}
\]
Also we define $x_{N+l}=-x_l$ for later use.

So the Hamiltonian can be approximated as:

\[H \simeq -\vec{B}_{eff}^{(q)}\sum_{i=1}^{n}\vec{s}^{(q)}_i\]
\[+(\frac{1}{2}fqn\sum_{l=1}^{N}x_l^2 +a<\vec{s}^{(q)}>\sum_{i=1}^{n}\vec{p}^{(q)}-na<\vec{p}^{(q)}><\vec{s}^{(q)}>)\]

Here $\vec{B}_{eff}^{(q)}$ is the mean-field, the projection of $\vec{B}_{eff}^{(q)}$ in direction l suffices:
\[
B_{eff l}=fqx_l+B_l-a<p_l>, l=1,2,3,\dots, q
\]

The partition function of the system is:
\[
Z=\sum_{\vec{s}^{(q)}_1}\dots \sum_{\vec{s}^{(q)}_n}e^{-\beta H}
\]
\[
\simeq e^{-\beta (\frac{1}{2}fqn\sum_{l=1}^{N}x_l^2+a<\vec{s}^{(q)}>\sum_{i=1}^{n}\vec{p}^{(q)}-na<\vec{p}^{(q)}><\vec{s}^{(q)}>)}
\sum_{\vec{s}^{(q)}_1}\dots \sum_{\vec{s}^{(q)}_n}e^{\beta \vec{B}_{eff}^{(q)}\vec{s}^{(q)}_i}
\]

since M of the n lattices is occupied by spins,
\[
Z
\simeq e^{-\beta (\frac{1}{2}fqn\sum_{l=1}^{N}x_l^2+a<\vec{s}^{(q)}>\sum_{i=1}^{n}\vec{p}^{(q)}-na<\vec{p}^{(q)}><\vec{s}^{(q)}>)}
A_{n}^{M}(\sum_{\vec{s}^{(q)}_i \in \textbf{S}\backslash \{\vec{e}_0\}} e^{\beta \vec{B}_{eff}^{(q)}\vec{s}^{(q)}_i})^{M}
\]

\subsubsection{Free energy}
Free energy of the system is;

\[
F=-\frac{1}{\beta}lnZ
\]
\[
\simeq (\frac{1}{2}fqn\sum_{l=1}^{N}x_l^2+a<\vec{s}^{(q)}>\sum_{i=1}^{n}\vec{p}^{(q)}-na<\vec{p}^{(q)}><\vec{s}^{(q)}>)
\]
\[
-\frac{1}{\beta}ln(A_{n}^{M})-\frac{1}{\beta}Mln(\sum_{\vec{s}^{(q)}_i \in \textbf{S}\backslash \{\vec{e}_0\}} e^{\beta \vec{B}_{eff}^{(q)}\vec{s}^{(q)}_i})
\]

\subsubsection{Details of Self-consistent Equations}

\textbf{1.High Temperature Limit Situation}

$\beta (fqx_l+B_l)<<1$ , so we can do Taylor expansion to the item $e^{\beta (fqx_l+B_l)}$

remember
$x_l=(<s_l>-<s_{l+N}>)=-x_{N+l}$and $(B_{N+1},B_{N+2},B_{N+3},\dots,B_{2N})=-(B_1,B_2,B_3,\dots,B_{N})$, so

\[n<s_l>\simeq M\frac{1+\beta (fqx_l+B_l)}{2N+\sum_{i=1}^{q}  \frac{1}{2}\beta^2 (fqx_i+B_i)^2}\]

so for $l=1$

\[n<s_1>\simeq M\frac{ 1+\beta (fqx_1+B_0)}{2N+\sum_{i=1}^{q}  \frac{1}{2}\beta^2 (fqx_i+B_i)^2}\]

\[n<s_{N+1}>\simeq M\frac{1+\beta (fqx_{N+1}-B_0)}{2N+\sum_{i=1}^{q}  \frac{1}{2}\beta^2 (fqx_i+B_i)^2}\]

for other $l$
\[n<s_l>\simeq M\frac{1+\beta fqx_l}{2N+\sum_{i=1}^{q}  \frac{1}{2}\beta^2 (fqx_i+B_i)^2}\]

\[n<s_{N+l}>\simeq M\frac{1+\beta fqx_{N+l}}{2N+\sum_{i=1}^{q}  \frac{1}{2}\beta^2 (fqx_i+B_i)^2}\]

then
\[
n(<s_1>-<s_{N+1}>)\simeq M\frac{\beta (fq(x_1-x_{N+1})+2B_0)}{2N+\sum_{i=1}^{q}  \frac{1}{2}\beta^2 (fqx_i+B_i)^2}
\]

\[
n(<s_l>-<s_{N+l}>)\simeq M\frac{\beta fq(x_l-x_{N+l})}{2N+\sum_{i=1}^{q}  \frac{1}{2}\beta^2 (fqx_i+B_i)^2}
\]

$x_l=-x_{N+l}=<s_l>-<s_{l+N}>$

\[
nx_1\simeq M\frac{2\beta (fqx_1+B_0)}{2N+\sum_{i=1}^{q}  \frac{1}{2}\beta^2 (fqx_i+B_i)^2}\]

\[
nx_l\simeq M\frac{2\beta fqx_l}{2N+\sum_{i=1}^{q}  \frac{1}{2}\beta^2 (fqx_i+B_i)^2}\]

then

\[
x_1\simeq \frac{\beta B_0}{N\frac{n}{M}-\beta fq}
\]

\[
x_l\simeq0, l=2,3,\dots,N
\]

\textbf{2.Low Temperature Limit Situation}

$\beta (fqx_l+B_l)>>1,\mbox{ if }(fqx_l+B_l)\neq 0$

\[n<s_l>=M\frac{e^{\beta (fqx_l+B_l)}}{\sum_{i=1}^{q} e^{\beta (fqx_l+B_l)}}\]

Guess the solution satisfies
\[
x_1>0,\mbox{ }
x_l=0,l=2,3,\dots,N
\]

\[n<s_l>=M\frac{e^{\beta (fqx_l+B_l)}}{2N-2+e^{\beta (fqx_1+B_0)}+e^{-\beta (fqx_1+B_0)}}\]

for $l=1$
\[n<s_1>=M\frac{e^{\beta (fqx_1+B_0)}}{2N-2+e^{\beta (fqx_1+B_0)}+e^{-\beta (fqx_1+B_0)}}=M, \mbox{ }\beta (fqx_1+B_1)>>1\]

the solution is
\[
<s_1>=\frac{M}{n}
\]
Easy to see that for others
\[
s_l=0, \mbox{ }l=2,3,\dots,2N
\]
so
\[
x_1=\frac{M}{n},\mbox{ } x_l=0,l=2,3,\dots,N
\]

which satisfies our guess.

\subsubsection{Details of self-organization}

\[\sum_{\vec{s}^{(q)}_i \in \textbf{S}\backslash \{\vec{e}_0\}} e^{-\beta \vec{B}_{eff}^{(q)}\vec{s}^{(q)}_i}=e^{\beta a<p_0>}\cdot \sum_{l=1}^{N}(e^{-\beta fqx_l }+e^{\beta fqx_l })
\]
\[\simeq 2e^{\beta a<p_0>}\cdot \sum_{l=1}^{N}( 1+\frac{1}{2}(\beta fqx_l)^2+\frac{1}{24}(\beta fqx_l)^4)\]

\[\simeq 2Ne^{\beta a<p_0>}\cdot ( 1+\sum_{l=1}^{N}\frac{1}{2N}(\beta fqx_l)^2+\sum_{l=1}^{N}\frac{1}{24N}(\beta fqx_l)^4)\]

\[ln(\sum_{\vec{s}^{(q)}_i \in \textbf{S}\backslash \{\vec{e}_0\}} e^{-\beta \vec{B}_{eff}^{(q)}\vec{s}^{(q)}_i})\simeq \]
\[ln(2N)+\beta a<p_0>+[\sum_{l=1}^{N}\frac{1}{2N}(\beta fqx_l)^2+\sum_{l=1}^{N}\frac{1}{24N}(\beta fqx_l)^4]
\]
\[
-\frac{1}{2}[\sum_{l=1}^{N}\frac{1}{2N}(\beta fqx_l)^2+\sum_{l=1}^{N}\frac{1}{24N}(\beta fqx_l)^4]^2
+\frac{1}{3}[\sum_{l=1}^{N}\frac{1}{2N}(\beta fqx_l)^2+\sum_{l=1}^{N}\frac{1}{24N}(\beta fqx_l)^4]^3\]

\[
\simeq ln(2N)+\beta a<p_0>+\sum_{l=1}^{N}\frac{1}{2N}(\beta fqx_l)^2+\sum_{l=1}^{N}(\frac{1}{24N})(\beta fqx_l)^4-\sum_{l=1}^{N}\sum_{k=1}^{N}\frac{1}{8N^2}(\beta fq)^4x_l^2x_k^2\]
\[-\sum_{l=1}^{N}\sum_{k=1}^{N}\frac{1}{96N^2}(\beta fq) ^6x_l^2x_k^4+
\sum_{l=1}^{N}\sum_{k=1}^{N}\sum_{m=1}^{N}(\frac{1}{24N^3})(\beta fq)^6 x_l^2x_k^2 x_m^2\]

\[
F\simeq -\frac{1}{\beta}ln(A_{n}^{M})-\frac{M}{\beta}\{ln(2N)+\beta a<p_0>\}\]

\[+(\frac{1}{2}fqn-\frac{\beta Mf^2q^2}{2N})\sum_{l=1}^{N}x_l^2-\frac{M}{\beta}\sum_{l=1}^{N}(\frac{1}{24N})(\beta fqx_l)^4+\frac{M}{\beta}\sum_{l=1}^{N}\sum_{k=1}^{N}\frac{1}{8N^2}(\beta fq)^4x_l^2x_k^2\]

\[+\frac{M}{\beta}\sum_{l=1}^{N}\sum_{k=1}^{N}\frac{1}{96N^2}(\beta fq)^6x_l^2x_k^4-\frac{M}{\beta}
\sum_{l=1}^{N}\sum_{k=1}^{N}\sum_{m=1}^{N}(\frac{1}{24N^3})(\beta fq)^6 x_l^2x_k^2 x_m^2\]

\[
\frac{dF}{dx_l^{2}}=(\frac{1}{2}fqn-\frac{M\beta f^2q^2}{2N} )+\frac{M\beta^3 f^4q^4}{4N}(\frac{1}{N}-\frac{1}{3})x_l^2+\frac{M\beta^3 f^4q^4}{4N^2}\sum_{k=1,k\not=l}^{N}x_k^2+\frac{M\beta^5 f^6q^6}{96N^2}\sum_{k=1}^{N}x_k^4\]

\[+\frac{M\beta^5 f^6q^6}{8N^2}(\frac{1}{6}-\frac{1}{N})\sum_{k=1}^{N}\sum_{m=1}^{N}x_k^2x_m^2
\]

\[
\frac{d^2F}{d(x_l^{2})^2}=\frac{M\beta^3 f^4q^4}{4N}(\frac{1}{N}-\frac{1}{3})+\frac{M\beta^5 f^6q^6}{48N^2}x_l^2+\frac{M\beta^5 f^6q^6}{4N^2}(\frac{1}{6}-\frac{1}{N})\sum_{k=1}^{N}x_k^2
\]

\[
\frac{d^2F}{dx_l^{2}dx_k^2}=\frac{M\beta^3 f^4q^4}{4N^2}+\frac{M\beta^5 f^6q^6}{48N^2}x_k^2+\frac{M\beta^5 f^6q^6}{4N^2}(\frac{1}{6}-\frac{1}{N})\sum_{b=1}^{N}x_b^2
\]

\textbf{When $\beta>\frac{nN}{Mfq}$}

From
$\frac{dF}{dx_l^2}=0$
we have

\[ \left( \begin{array}{ccccccc}
1-\frac{N}{3} & 1 & \dots & 1 & 1 & 1 \\
1 & 1-\frac{N}{3} & \dots & 1 & 1 & 1\\
\dots & \dots & \dots & \dots & \dots & \dots\\
1 & 1 & \dots & 1 & 1-\frac{N}{3} & 1\\
1 & 1 & \dots & 1 & 1 & 1-\frac{N}{3}\\
 \end{array}\right)\left( \begin{array}{ccccccc}
x_1^2 \\
x_2^2 \\
\dots \\
x_{N-1}^2 \\
x_N^2 \\
\end{array}\right)\]

\[
=\frac{4N^2}{M\beta^3 f^4q^4}(\frac{M\beta f^2q^2}{2N}-\frac{fqn}{2})\left(\begin{array}{c}
1\\
1\\
\dots\\
1\\
1\\
 \end{array}\right)\]

so
\[x_l^2=\frac{3}{\beta^2f^2q^2}(1-\frac{Nn}{M\beta fq})\]

or

\[x_l=\pm\sqrt{\frac{3}{\beta^2f^2q^2}(1-\frac{Nn}{M\beta fq})}\]

\end{widetext}

\end{document}